\documentclass[submission, Phys]{SciPost}

\def\pinit{p^\ast}
\def\qinit{q^\ast}

\usepackage{amsmath}
\usepackage{amsfonts}
\usepackage{amssymb}
\usepackage{accents}
\usepackage{bm}
\usepackage{color}
\usepackage{graphicx}
\usepackage{amsthm}
\usepackage{enumerate}
\usepackage{mathrsfs}
\newlength{\dhatheight}
\newcommand{\doublehat}[1]{%
    \settoheight{\dhatheight}{\ensuremath{\hat{#1}}}%
    \addtolength{\dhatheight}{-0.35ex}%
    \hat{\vphantom{\rule{1pt}{\dhatheight}}%
    \smash{\hat{#1}}}}

\begin{document}


\begin{center}{\Large \textbf{Extreme value statistics of edge currents in Markov jump processes and their use for entropy production estimation  }}\end{center}

\begin{center}
Izaak Neri\textsuperscript{1}, Matteo Polettini\textsuperscript{2}
\end{center}

\begin{center}
{\bf 1} Department of Mathematics, King’s College London, Strand, London, WC2R 2LS, UK
izaak.neri@kcl.ac.uk
\end{center}

\begin{center}
{\bf 2} Department of Physics and Materials Science, University of Luxembourg, Campus Limpertsberg, 162a avenue de la Fa\"iencerie, L-1511 Luxembourg (G. D. Luxembourg)
matteo.polettini@uni.lu
\end{center}

\section*{Abstract}
{\bf
 The infimum of an integrated  current is its extreme value against the direction of its average flow.    Using martingale theory, we show that  the infima of integrated edge currents in  time-homogeneous Markov jump processes are geometrically distributed, with    a mean value determined by the effective affinity measured by  a marginal observer that only sees the integrated edge current.   In addition, we show that a marginal observer  can estimate a finite fraction of the average  entropy production rate in the underlying  nonequilibrium process from the extreme value statistics in the  integrated edge current.   The estimated average  rate of dissipation obtained in this way equals the 
  above mentioned effective affinity times the average edge current.   Moreover, we show that estimates of dissipation based  on extreme value statistics can be significantly more accurate than those based on thermodynamic uncertainty ratios, as well as  those based on  a naive estimator obtained by   neglecting  nonMarkovian correlations in  the Kullback-Leibler divergence of the  trajectories of the integrated edge current.
}  

\vspace{10pt}
\noindent\rule{\textwidth}{1pt}
\tableofcontents\thispagestyle{fancy}
\noindent\rule{\textwidth}{1pt}
\vspace{10pt}

\section{Introduction}  
Currents with nonzero average value are a hallmark of   nonequilibrium processes.   In statistical physics, there has been much interest in  characterising  the statistics of currents, with initial work focusing on fluctuation relations~\cite{andrieux2007fluctuationa, andrieux2007fluctuation, gaspard2013multivariate}.     More recently, it was shown that the large deviation rate  function of a current is upper bounded by a parabola with a prefactor that is proportional to the entropy production rate  \cite{pietzonka2015universal, gingrich2015dissipation} and that the Fano factor of currents is bounded from below by the inverse dissipation rate \cite{barato2015thermodynamic, PhysRevE.96.012101,  PhysRevE.96.020103}.    Hitherto,  current fluctuations have mainly been   considered at  fixed times.   However, since  currents are stochastic processes, it is possible to quantify current fluctuations through other  properties of a trajectory, such as  the first-passage properties  \cite{roldan2015decision, saito2016waiting, PhysRevE.95.032134, gringich2017bis,  PhysRevE.103.L050103, PhysRevResearch.3.L032034, neri2021universal, wampler2021skewness, PhysRevE.105.044127, neri2} or the (closely related) extreme value statistics of currents~\cite{hartich2019extreme}.

  \begin{figure}[t!]\centering
 \includegraphics[width=0.5\textwidth]{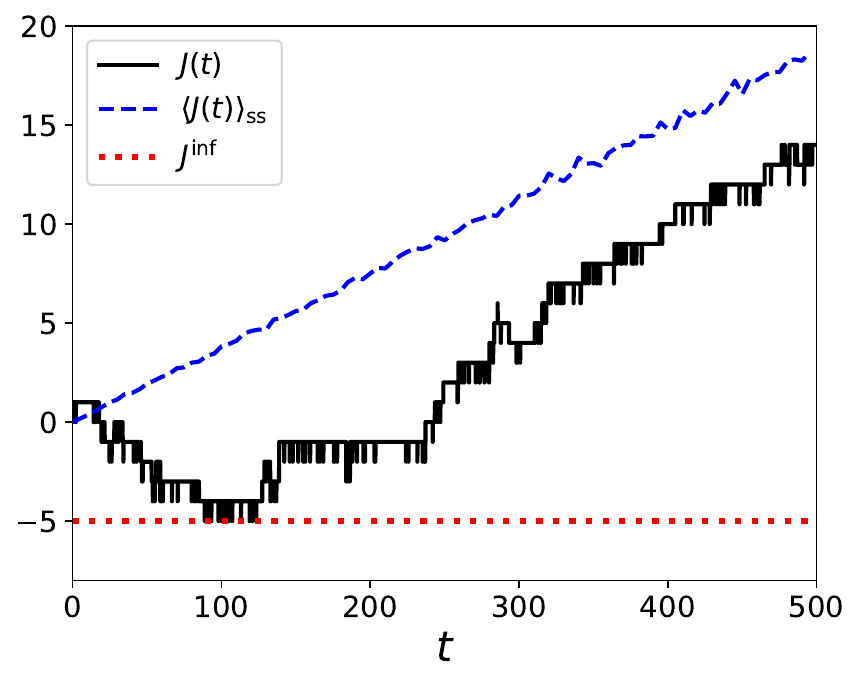}
\caption{Illustration of the infimum $J^{\rm inf}$ (dotted, red line) in a trajectory $J(t)$ (black, solid line) of an empirical time-integrated current  with positive mean flow $\langle J(t)\rangle_{\rm ss}$ (blue, dashed line).   The data is taken from a  
trajectory of the position variable $J=J_{\rm pos}$ in the kinesin-1 model defined in Sec.~\ref{sec:kinesin} with parameters $[{\rm ATP}] = 0.1\mu{\rm M}$ and $f_{\rm mech} = 3.5{\rm pN}$. } \label{fig:infIl}
\end{figure} 

 In the present paper we focus on this latter.  
Consider an empirical time-integrated current    $J(t)$ in a nonequilibrium stationary state of a classical, stochastic process.  The time index $t\in\mathbb{R}^+$,  and we  use the conventions that the current is zero at the origin of time  and increases on average.   The infimum $J^{\rm inf}$ of the current, as illustrated by the red dotted line in Fig.~\ref{fig:infIl}, is the most negative value that the current takes, and hence it determines its largest excursion  in the direction that opposes  the average flow.    

Adding to the fact that extreme value  statistics of stochastic processes are interesting in their own right, see e.g.~Ref.~\cite{majumdar2020extreme},  
there are also a couple of specific reasons from  nonequilibrium thermodynamics why we would like to study extreme values of currents.
A first reason is because  the statistics of   extreme values  exhibit universal properties that are  analytically tractable with martingale methods~\cite{chetrite2011two, neri2017statistics, neri2019integral, PhysRevE.105.024112}.   For example, for the  entropy production $S$, which is one the most  well-studied examples of a current,  exact results have been derived for the statistics of the  infimum value $S^{\rm inf}$.  References~\cite{neri2017statistics, neri2019integral} show that the mean infimum  of the entropy production is greater than or equal to one,  i.e.,
\begin{equation}
\langle S^{\rm inf} \rangle_{\rm ss}\geq-1,  \label{eq:1}
\end{equation}
and an analogous bound holds for the cumulative distribution of the infimum \cite{neri2017statistics, neri2019integral}, viz., 
\begin{equation}
\mathbb{P}_{\rm ss}\left(S^{\rm inf}\leq    -\ell\right) \leq e^{-\ell} ,    \quad  \forall \ell\in \mathbb{R}^+,  \label{eq:2}
\end{equation}
where $\mathbb{P}_{\rm ss}\left(\cdot\right)$ is the probability measure in the stationary state and $\langle \cdot \rangle_{\rm ss}$ denotes the average with respect of $\mathbb{P}_{\rm ss}$;  when  $S(t)$ is continuous the equalities in Eqs.~(\ref{eq:1})-(\ref{eq:2})   are attained.   Equations (\ref{eq:1})-(\ref{eq:2})    constrain  negative fluctuations of the entropy production, which have also been studied in experimental setups \cite{singh2019extreme,  PhysRevE.102.062127}.

Note that the infimum statistics of entropy production, as given by Eqs.~(\ref{eq:1})-(\ref{eq:2}), follow from the fact that  $e^{-S}$  is a martingale.   This latter property is a direct consequence of the fact that $e^{-S}$ is a density process --- also known as the Radon-Nikodym derivative process \cite{liptser1977statistics, neri2017statistics, PhysRevE.101.022129} --- relating the statistics of two probability measures, viz., the probability measure  $\tilde{\mathbb{P}}_{\rm ss}$ of the time-reversed process and the probability measure $\mathbb{P}_{\rm ss}$ of the forward process.    It should be noted that  Radon-Nikodym derivative processes of  arbitrary measures with respect to $\mathbb{P}_{\rm ss}$ are martingales, see Appendix~\ref{App:A}.  Hence the martingale approach can be extended to  processes other than the entropy production,  as long as an appropriate probability measure can be identified, see e.g,.~Refs.~\cite{chetrite2011two, chetrite2019martingale, roldan2022martingales}.

A second reason why extreme values of currents are interesting is because  they can be used to estimate the average rate $\dot{s}$  in stationary  processes consisting of variables that have even parity under time reversal, which is the case when inertia is negligible and external forces are not governed by magnetic fields,     see Refs.~\cite{roldan2015decision, neri2021universal, neri2}.  
Indeed, let $J(t)$ be an  integrated, empirical current  in an overdamped, Langevin process or a Markov jump process,  and let $\overline{j} = \langle J(t)\rangle_{\rm ss}/t>0$ be its rate on average.   If we define 
\begin{equation}
    \hat{s}_{\rm inf}(\ell) := \frac{\left|\ln p_-(\ell)\right|}{\ell}\overline{j} , \label{eq:sinf}
\end{equation}
where 
\begin{equation}
p_-(\ell) := \mathbb{P}_{\rm ss}\left(J^{\rm inf}\leq -\ell\right)
\end{equation}
is the  probability that  the infimum value  $J^{\rm inf}$  is smaller or equal than $-\ell$, then
\begin{equation}
    \lim_{\ell\rightarrow \infty}    \hat{s}_{\rm inf}(\ell) \leq \dot{s}. \label{eq:sinf2}
\end{equation}
The equality here is attained when $J$ is proportional to $S$.     
Note that although the relations Eqs.~(\ref{eq:sinf}) and (\ref{eq:sinf2}) did not appear before in the literature, they can be seen as a particular case of the first-passage ratio $\hat{s}_{\rm FPR}$ in Ref.~\cite{neri2} when the positive threshold of the associated first-passage problem diverges.     Interestingly, far from  equilibrium, $\hat{s}_{\rm inf}(\ell)$, for large values of $\ell$, captures a finite fraction of the total rate of entropy production, while estimates based on the variance of the current capture a negligible fraction of the total entropy \cite{neri2}.    Hence, far from  equilibrium, it is  more effective to use extreme values of currents to estimate the rate $\dot{s}$ of entropy production  than to use the variance of the current.

In the present paper, we study in detail the  infima statistics of empirical, integrated, {\it edge currents} in stationary, Markov jump processes, i.e., currents along the edges of a graph representing the different possible transitions in state space.   Following Refs.~\cite{polettini2017effective, polettini2019effective}, we obtain a martingale that is closely related to the edge currents of a Markov jump process  and that is the Radon-Nikodym derivative process of a measure, which we call $\mathbb{R}_{\rm ss}$, with respect to $\mathbb{P}_{\rm ss}$.      Subsequently  we use martingale  manipulations, similar to those presented in Ref.~\cite{neri2019integral}  for the  exponentiated negative entropy production, to determine the statistics of extreme values of  edge currents.

In stationary  processes with variables that have even parity under time reversal, the derived results for the extreme value statistics of edge currents can be interpreted in terms of an effective thermodynamic picture of a  {\it marginal observer} that only sees the edge current $J$, ignorant of the existence of other currents in the system.   Such an observer  assumes that $J$ is proportional to the entropy production and  measures an effective affinity $a(t)$ through the relation \cite{PhysRevLett.117.180601, polettini2017effective, polettini2019effective}
\begin{equation}
    \langle e^{-a(t) J(t)}\rangle_{\rm ss}  =1   . \label{eq:integralA}
\end{equation}
In this paper we show that: (i) the  affinity  
\begin{equation}
\lim_{t\rightarrow \infty}a(t)=a^\ast
\end{equation}
determines   the  extreme value statistics of the edge current; (ii)   using extreme value statistics, a marginal observer estimates a dissipation rate $\overline{j}a^\ast$,
i.e., we identify     
\begin{equation}
\lim_{\ell\rightarrow \infty}\frac{|\ln p_-(\ell)|}{\ell} = a^\ast, \label{eq:asta}
\end{equation}
connecting the thermodynamics of the edge current (according to Eqs.~(\ref{eq:sinf}), (\ref{eq:sinf2}), and (\ref{eq:asta}))  with its kinematics (according to Eq.~(\ref{eq:integralA})).  These results  show, in line with the results in Refs.~\cite{PhysRevLett.117.180601, polettini2017effective,bisker2017hierarchical,  polettini2019effective}, that  although the edge current $J$ is non-Markovian, a marginal observer can develop a consistent, effective  thermodynamics.

According to Eq.~(\ref{eq:sinf2}), estimating $\dot{s}$ with 
$\hat{s}_{\rm inf}(\ell)$ requires measurements of the cumulative probability $p_-(\ell)$ at large thresholds $\ell$.   This is  undesirable as the probability $p_-(\ell)$  decays exponentially fast as a function of $\ell$, and therefore it is difficult to empirically estimate $p_-$ at large values of $\ell$; we call this the {\it infinite threshold} problem \cite{roldan2015decision, neri2021universal,neri2}.  In this paper, for the particular case of  integrated edge currents, i.e., $J = J_{x\rightarrow y}$, we resolve the infinite threshold problem with the estimator
\begin{equation}
\doublehat{s}_{\rm inf}(\ell)    :=  \overline{j}  \ln \frac{p_-(\ell)}{p_-(\ell+1)} ,  \label{eq:modified}
\end{equation}
for which we show that 
\begin{equation}
\doublehat{s}_{\rm inf}(\ell)  = \lim_{\ell'\rightarrow \infty}\hat{s}_{\rm inf}(\ell')  =  \overline{j}a^\ast \leq \dot{s}    \label{eq:finiteA} 
\end{equation}
 for all $\ell\in \mathbb{N}$.   Hence, using the estimator $\doublehat{s}_{\rm inf}(\ell)$ for $\ell\neq 0$, a proportion of the  average rate  of dissipation, $\dot{s}$,  can be estimated from the  probability $p_-(\ell)$ that the infimum of an edge current is smaller than  a  {\it finite} threshold value $\ell$, resolving the infinite threshold problem.        
 
 The quantity $\overline{j}a^\ast$ has also been studied in Ref.~\cite{bisker2017hierarchical}, where it is called the average  {\it informed partial} entropy production rate.   Reference~\cite{bisker2017hierarchical}   shows that   $\overline{j}a^\ast$ is a better estimate of dissipation than a naive estimator $\hat{s}_{\rm KL}$ obtained from   neglecting  nonMarkovian correlations in  the Kullback-Leibler divergence of the  trajectories of the current, and which is called the average {\it passive partial} entropy production rate, i.e.,  $\dot{s} \geq \overline{j}a^\ast \geq \hat{s}_{\rm KL} $.   However,  Reference~\cite{bisker2017hierarchical}     argues that  $\overline{j}a^\ast$ cannot be measured passively by observing the trajectory of a current, and instead should be determined as the force at which the edge current stalls, which can be measured actively if we can exert a microscopic force on the system.   In the present paper, we show that  the informed partial entropy production rate,       $\overline{j}a^\ast$, can be measured passively through the modified infimum ratio $\doublehat{s}_{\rm inf}(\ell)$, as $\doublehat{s}_{\rm inf}(\ell) = \overline{j}a^\ast$ for $\ell\in \mathbb{N}$.

The paper is organised as follows:
We summarise  in Sec.~\ref{sec:3x} the main results.    Before addressing the general problem,  we derive in Sec.~\ref{sec:2x} the infimum statistics of  an edge current that is proportional to the  entropy production; this is a special  case that is easily solvable and gives  an idea of the results we obtain and  mathematical methods we use in the general case. 
In Sec.~\ref{sec:II}, we introduce the system setup and some of the mathematical groundwork that we use in later sections to derive  the main results.  In Sec.~\ref{sec:III}, following Refs.~\cite{polettini2017effective, polettini2019effective}, we  introduce a set of martingales associated with the empirical, integrated, edge currents of  stationary, Markov jump processes, which constitute the main mathematical tool that  permits us to obtain the main results.    Subsequently, in Sec.~\ref{sec:IV}, we use the concepts from Secs.~\ref{sec:II} and \ref{sec:III} to derive the main result, which is an explicit expression for the probability mass function of the infima of empirical, integrated, edge currents.     In Sec.~\ref{sec:kinesin}, we show the main result at work on a simple model of two-headed molecular motors~\cite{Liepelt, hwang2018energetic, hwang2019correction}.  In Sec.~\ref{sec:appl}, we study the properties of  the estimators $\hat{s}_{\rm inf}$ and $\doublehat{s}_{\rm inf}$ for the average rate $\dot{s}$ of dissipation based on the infimum statistics of empirical, integrated, edge currents.    We end the paper with a discussion in Sec.~\ref{sec:VII}.   The paper also contains a few appendices with details about some of the derivations and the model defined in Sec.~\ref{sec:kinesin}.

\section{Summary of the main results}\label{sec:3x}

We first summarise the main results, and then we discuss how these results are related to the companion paper Ref.~\cite{companion}.  

\subsection{Main results}\label{sec:31}

 Let $X(t)\in \mathcal{X}$ be a time-homogeneous Markov jump process, and let 
\begin{equation}
J_{x\rightarrow y}(t) := N_{x\rightarrow y}(t)-N_{y\rightarrow x}(t), \quad {\rm with} \quad x,y\in \mathcal{X}, \label{eq:JEdge}
\end{equation}
  denote the difference between the number of times $N_{x\rightarrow y}(t)$  that $X$ has jumped from $x$ to $y$ in the time interval $[0,t]$ minus the number of times $N_{y\rightarrow x}(t)$ that  $X$  has jumped from $y$ to $x$ in the same interval of time.    Let us assume, without loss of generality, that $\langle J_{x\rightarrow y}(t)  \rangle_{\rm ss} >0$ when $t$ is large enough.  
  
 We call $J_{x\rightarrow y}(t)$ an empirical, integrated {\it edge current}, as it is the flow along the edge $x\rightarrow y$ of the graph of possible transitions in the phase space  $\mathcal{X}$; we call $x$ the {\it source} node and $y$ the target node of the edge current.   Notice that, for convenience, we often speak of edge currents, tout court, and  it should be understood that we consider empirical, integrated currents.      Edge currents are the elementary currents of a Markov jump process.  Indeed,  empirical, time-integrated currents  $J$ can be expressed as a  linear combination 
\begin{equation}
J := \sum_{(u,v) \in \mathcal{E}}c_{u,v} J_{u\rightarrow v}(t)   \label{eq:JGeneral}  
\end{equation}
of the  edge currents, where $c_{u,v}$ is the amount of a certain resource that is exchanged or transported to/from the environment when the process jumps from $u$ to $v$, and where $\mathcal{E}\subset\mathcal{X}^2$ is the  set of pairs $(u,v)$ with nonzero transition rates; note that we consider reversible processes for which $(u,v)\in \mathcal{E}\Leftrightarrow (v,u)\in \mathcal{E}$.

 The fluctuations of  the edge current $J_{x\rightarrow y}$ against the direction of the  average flow can be characterised by the infimum 
 \begin{equation}
J^{\rm inf}_{x\rightarrow y} := {\rm inf}_{t\geq  0} J_{x\rightarrow y}(t).
\end{equation}  
Note that $J_{x\rightarrow y}^{\rm inf}$ is a nonpositive integer as $J_{x\rightarrow y}(0)=0$.    If    $\langle J_{x\rightarrow y}(t)  \rangle_{\rm ss} <0$, then we can consider the infimum of $-J_{x \rightarrow y}$, which is the supremum of $J_{x \rightarrow y}$.  

In this Paper, using martingale methods similar to those used  in Refs.~\cite{neri2017statistics, neri2019integral} to derive the infimum law Eq.~(\ref{eq:1}), we show that
the probability mass function of  $J_{x\rightarrow y}^{\rm inf}$ is given by
\begin{equation}
p_{J_{x\rightarrow y}^{\rm inf}}(-\ell|X(0)=x_0) =   \left\{\begin{array}{ccc} e^{-\ell a^\ast_{x\rightarrow y}}(1-p_{\rm esc}(x_0)) (e^{a^\ast_{x\rightarrow y}}-1), &{\rm if}& \ell \in \mathbb{N}, \\ p_{\rm esc}(x_0),&{\rm if}& \ell=0,  \end{array}\right.\label{eq:distriExp}
\end{equation}
and its mean value by
\begin{equation}
\langle J_{x\rightarrow y}^{\rm inf}|X(0)=x_0\rangle_{\rm ss}  = -\frac{1-p_{\rm esc}(x_0)}{1-e^{-a^\ast_{x\rightarrow y}}}, \label{eq:Infx}
\end{equation} 
where $a^\ast_{x\rightarrow y}>0$ is an ``effective'' affinity that was identified before in Refs.~\cite{PhysRevLett.117.180601, polettini2017effective, bisker2017hierarchical, polettini2019effective}, and $p_{\rm esc}(x_0)$ is the probability that the infimum equals zero. In Eqs.~(\ref{eq:distriExp}) and (\ref{eq:Infx}) we have used  probabilities and expectation values conditioned on a general, initial state $X(0)=x_0$.       

In general,  $p_{\rm esc}(x_0)$ does not admit a simple expression in terms of $a^\ast_{x\rightarrow y}$.   A notable exception is when $X(0)=x$, where $x$ is the source node of the edge $x\rightarrow y$, in which case 
\begin{equation}
    p_{\rm esc}(x) =1-e^{-a^\ast_{x\rightarrow y}}.\label{eq:simpleCase}
\end{equation}

When $|a^\ast_{x\rightarrow y}|\ll 1$  the current   is stalled at an almost zero average rate, i.e., $\langle J_{x\rightarrow y}(t)\rangle_{\rm ss}\approx 0$,  and consequently $p_{\rm esc}(x_0)\approx 0$, such that    the geometric distribution Eq.~(\ref{eq:distriExp}) is approximately an exponential distribution.   Equilibrium states are examples of stalled states, but it is also possible to have  stalled currents far from  equilibrium.  Notably, a marginal observer that only measures $J_{x\rightarrow y}(t)$ cannot distinguish between an equilibrium state and a nonequilibrium stalled state from the measurements of extreme values of $J_{x\rightarrow y}(t)$.

Equation~(\ref{eq:distriExp}) implies that the fluctuations of the extreme values of $J_{x\rightarrow y}(t)$ are determined by the effective affinity  $a^\ast_{x\rightarrow y}$.   
The effective affinity  $a^\ast_{x\rightarrow y}$ can be defined through the integral fluctuation relation~\cite{PhysRevLett.117.180601, polettini2017effective, polettini2019effective}
\begin{equation}
    \lim_{t \rightarrow \infty}\langle e^{-a^\ast_{x\rightarrow y}J_{x\rightarrow y}(t)}\rangle_{\rm ss} = 1  \label{eq:astJx}
\end{equation}
and hence admits a kinematic interpretation.  Indeed, applying Jensen's inequality  we find
\begin{equation}
    a^\ast_{x\rightarrow y} \langle J_{x\rightarrow y}(t)\rangle_{\rm ss}  \geq  0.
\end{equation}
In addition, assuming that the rates $k_{x \to y}$ and $k_{y \to x}$, governing the current of interest, are tunable, then the effective affinity is the difference of their log-ratio $\ln k_{x \to y} / k_{y \to x}$ to values where the average stationary current stalls, $\langle J_{x \to y} \rangle_{\rm ss} = 0$~\cite{polettini2017effective, polettini2019effective}.  In systems where these rates are regulated by large reservoirs of energy, particles, or (even) information, the effective affinity is the difference of the relevant thermodynamic potentials from the value where the system attains the stalling state.  
For this reason,  Ref.~\cite{bisker2017hierarchical} calls  $\langle J_{x\rightarrow y}(t)\rangle a^\ast_{x\rightarrow y}$    the informed partial entropy production.

As a last result, using Eq.~(\ref{eq:distriExp}), we show that $a^\ast_{x\rightarrow y}$ has a  thermodynamic meaning.   In particular, we show that $a^\ast_{x\rightarrow y}$ determines the average entropy production rate that a  marginal observer estimates from the measurement of the trajectories of   $J_{x\rightarrow y}$.    Indeed, substitution of  Eq.~(\ref{eq:distriExp}) in the estimator $\hat{s}_{\rm inf}$ of $\dot{s}$ --- defined in Eq.~(\ref{eq:sinf}) ---   we obtain that 
\begin{equation}
   \lim_{\ell \rightarrow \infty}  \hat{s}_{\rm inf}(\ell)  =   \overline{j}_{x\rightarrow y} a^\ast_{x\rightarrow y}, \label{eq:FPREst}
\end{equation}
where 
\begin{equation}
\overline{j}_{x\rightarrow y} := \langle J_{x\rightarrow y}(t)\rangle_{\rm ss}/t \label{eq:edgeCurrentDef}
\end{equation}
is the average current.   Hence, according to Eq.~(\ref{eq:FPREst}),  $a^\ast_{x\rightarrow y}$ is an effective thermodynamic affinity that when multiplied with the average current rate $\overline{j}_{x\rightarrow y}$ determines the entropy production rate $\hat{s}_{\rm inf}$ measured by a marginal observer.   Equation~(\ref{eq:sinf2}) together with Eq.~(\ref{eq:FPREst}) implies 
\begin{equation}
    \overline{j}_{x\rightarrow y}a^\ast_{x\rightarrow y} \leq \dot{s},
\end{equation}
an inequality that has also been derived directly from  the properties of the  generator of the underlying Markov process, as shown in Ref.~\cite{polettini2019effective}.  

Equation~(\ref{eq:FPREst}) is an asymptotic result for large thresholds, as it is based  on the estimator $\hat{s}_{\rm inf}(\ell)$ in the limit of large $\ell$.   
However,  from Eq.~(\ref{eq:distriExp}) it follows that  the effective affinity can be estimated using
\begin{equation}
    a^\ast_{x\rightarrow y} =
    \ln \frac{p_-(\ell)}{p_-(\ell+1)},
\end{equation}
for all $\ell\in \mathbb{N}$, and hence the effective affinity can be obtained from the measurement of the probability that an edge current goes below a certain {\it finite} threshold, thus resolving for the case of edge currents the infinite threshold problem~\cite{roldan2015decision, neri2021universal,neri2}.

\subsection{Relation to the companion paper Ref.~\cite{companion}}\label{sec:32}
The present manuscript comes with the companion manuscript Ref.~\cite{companion} that addresses similar questions.  Reference \cite{companion} focuses on the probability \begin{equation}
\mathfrak{f}_- = 1-p_{\rm esc}(x_0) = \mathbb{P}_{\rm ss}\left(J^{\rm inf}_{x\rightarrow y}\leq -1|X(0)\right)
\end{equation} 
that the infimum of  $J_{x\rightarrow y}$ is smaller  or equal than $-1$, instead of on its probability mass function.   However, the main difference between  Ref.~\cite{companion} and the present manuscript is from a methodological point of view.  Reference \cite{companion} derives the result Eq.~(\ref{eq:distriExp})  by identifying  a Markov process in transition space, while the present manuscript  identifies a martingale process associated with $J_{x\rightarrow y}$, and subsequently uses this martingale to derive Eq.~(\ref{eq:distriExp}).     Both     approaches have been developed independently, and consequently both manuscripts can be read independently.      Taken together,  we believe it is interesting to see how the exact solvability of this problem materialises into two different ways.  

Comparing both manuscripts, the following dictionary is useful: Ref.~\cite{companion} uses $\rho(v|u)$ for the transition rates $k_{u\rightarrow v}$, $1\rightarrow 2$ for the observed edge $x\rightarrow y$, $c$ for  integrated currents $J$, $F$ for the effective affinity $a^\ast_{x\rightarrow y}$, the subindex  $\infty$ for  stationary states instead of the subindex     ${\rm ss}$, $\mathfrak{p}_{-n}[p^\mathcal{L}_1]$ for the probability mass function $p_{J^{\mathrm{inf}}_{x \to y}} (-\ell | X(0) = x_0)$ of the infimum, and $\mathfrak{p}_{0}$ for the escape probability $p_{\rm esc}$.

\section{Prelude: extreme value statistics of edge currents that are proportional to the  entropy production}\label{sec:2x}
In this Section, as a simplified initial problem, we derive the statistics of infima of  edge currents that 
are proportional to the  entropy production.    This problem is relevant both from a mathematical and a physical point of view.   From  a mathematical point of view,   the statistics of  currents that are proportional to the entropy production can be determined readily from the fact that $e^{-S}$ is a martingale, see Refs.~\cite{neri2017statistics, neri2019integral}, and this  constitutes  an introduction in a simplified setup to the methods we will use  in this paper.   From  a physical  point of view,  a marginal observer that only observes the edge current $J_{x\rightarrow y}$, unaware of the existence of other currents in the system, thinks that the observed current $J_{x\rightarrow y}(t)$ is proportional to  the entropy production.   Hence, it is insightful to compare the  main result  Eq.~(\ref{eq:distriExp})  with the infimum statistics of the entropy production.

We consider  the  entropy production $S(t)$ of a nonequilibrium process that takes the form 
\begin{equation}
S(t) = c\: J_{x\rightarrow y}(t),
\end{equation}
where $c>0$ is a constant, proportionality factor, sometimes called  ``affinity" from pre-modern alchemic theories of the combination of elements, see Ref.~\cite{Goethe}.     

Consider  the stopping problem of establishing the first time entropy production exits a certain interval, i.e.,
\begin{equation}
T := {\rm inf}\left\{t\geq 0: S(t) \notin (-\ell_-c,\ell_+c)\right\},\quad {\rm with} \quad \ell_-,\ell_+\in\mathbb{N}. \label{eq:stoppingxx}\end{equation}

Since the interval $(-\ell_-c,\ell_+c)$ is finite, it holds that 
\begin{equation}
    p_- + p_+  =1, \label{eq:stop}
\end{equation}
where $p_-$ is the probability that both $T<\infty$ and $S(T) \leq  -\ell_- c$ hold, and $p_+$ is the probability that both $T<\infty$ and $S(T) \geq  \ell_+ c$ hold.    

In addition, since $e^{-S}$ is a martingale \cite{chetrite2011two, neri2017statistics, neri2019integral}, the integral fluctuation relation at stopping times  \cite{neri2019integral}
\begin{equation}
    \langle e^{-S(T)}\rangle_{\rm ss} = 1
\end{equation}
applies, and therefore  
\begin{equation}
    p_-  \langle e^{-S(T)} |S(T) \leq  -\ell_- c  \rangle_{\rm ss} +  p_+ \langle e^{-S(T)} |S(T) \geq  \ell_+ c  \rangle_{\rm ss} =1 .\label{eq:stoppingaxa}
\end{equation} 
Since $J_{x\rightarrow y}$ changes in increments of size $\pm 1$,  the entropy production $S =cJ_{x\rightarrow y}$ changes in discrete increments of $\pm c$, and since $S(0) = J_{x\rightarrow y}(0) = 0$,  Eq.~(\ref{eq:stoppingaxa}) yields
\begin{equation}
    p_-  e^{\ell_- c}  +  p_+ e^{-\ell_+ c} =1.  \label{eq:stopping2}
\end{equation} 

Combining the Eqs.~(\ref{eq:stop}) and (\ref{eq:stopping2}), we obtain  that
\begin{equation}
    p_- =\frac{1 -e^{-\ell_+ c}}{e^{\ell_- c}-e^{-\ell_+ c}} .  \label{eq:stopping2x}
\end{equation} 
In the limit of $\ell_+\rightarrow \infty$,  Eq.~(\ref{eq:stopping2x}) reduces to 
\begin{equation}
    p_- = e^{-\ell_- c} , \quad   \forall \ell_- \in \mathbb{N}.  
\end{equation}
Identifying 
\begin{equation}
    p_- = \mathbb{P}_{\rm ss}\left(S^{\rm inf}\leq -c\ell_-\right),
\end{equation}
we obtain for the probability mass function of $S^{\rm inf}$~\cite{guillet2020extreme}, 
\begin{equation}
p_{S^{\rm inf}}(-c\ell) = \mathbb{P}_{\rm ss}\left(S^{\rm inf} = -c\ell\right) =  e^{-\ell c}(1-e^{-c}), \quad \forall \ell \in \mathbb{N}\cup \left\{0\right\}.\label{eq:distriExpxax}
\end{equation}
The average infimum is thus 
\begin{equation}
\langle S^{\rm inf}\rangle_{\rm ss} = -\frac{c}{e^{c}-1}\geq -1, \label{eq:Infxxax}
\end{equation} 
consistent with the infimum law Eq.~(\ref{eq:1}).

Comparing  Eqs.~(\ref{eq:distriExp}) and (\ref{eq:distriExpxax}), we conclude that, ignoring  the prefactor $p_{\rm esc}(x_0)$ and the value of the probability mass function at $\ell=0$, a marginal observer measures a  statistics  for $J_{x\rightarrow y}^{\rm inf}$  that is equivalent to the statistics of the   entropy production in a system  for which   $S = a^\ast_{x\rightarrow y} J_{x\rightarrow y}$.      

In the following Section we define the system setup in which we will derive Eq.~(\ref{eq:distriExp}) in full generality.

\section{System setup and mathematical groundwork}  \label{sec:II}
We introduce the system setup and some of the  mathematical tools that we use to derive the main results.  
\subsection{Markov jump processes}
 Let $(\Omega, \mathscr{F})$ be  a measurable space, with $\Omega$ the set of realisations $\omega\in\Omega$ of a physical process, and $\mathscr{F}$ a $\sigma$-algebra of measurable events.     Let  $X(\omega, t) = X(t)$,  with $\omega\in \Omega$ and $t\in \mathbb{R}^+$  a continuous time index, be a stochastic process defined on $(\Omega, \mathscr{F})$ and that takes values in a finite set $\mathcal{X}\ni X(t)$.    Notice that the realisations $\omega$ consist of trajectories over the interval $t
 \in [0,\infty)$, that $X(\omega,t)$ returns the value of the trajectory at time $t$, and that the set of measurable events contains, amongst others, the sets $\left\{\omega\in \Omega: X(t,\omega)=x\right\}$ for all $t\geq 0$ and $x\in\mathcal{X}$.   We denote trajectories of $X$ over a finite interval $[0,t]$  by $X^t_0$.

 To determine the statistics of $X$, we consider a probability measure $\mathbb{P}_{\pinit}$ defined on  $(\Omega, \mathscr{F})$, where $\pinit$ is the probability distribution of the initial configuration $X(0)$.    Notice that  probability measures assign probabilities $\mathbb{P}_{\pinit}(\Phi)\in [0,1]$  to events $\Phi\in\mathscr{F}$ in the $\sigma$-algebra $\mathscr{F}$.    If we  observe trajectories $X^t_0$ in a fixed time interval $[0,t]$, then there is no need to consider the full $\sigma$-algebra $\mathscr{F}$ generated by infinitely long trajectories $X^{\infty}_0$.   Instead, it is then sufficient to consider the sub-$\sigma$-algebra $\mathscr{F}_t$ generated by trajectories $X^t_0$ over a finite time window, and we denote the measure $\mathbb{P}_{\pinit}$ restricted to  $\mathscr{F}_t$  by $\mathbb{P}_{\pinit}[X^t_0]$.   In other words, $\mathbb{P}_{\pinit}[X^t_0]$ is the probability measure defined on $\mathscr{F}_t$ such that  $\mathbb{P}_{\pinit}[X^t_0](\Phi) = \mathbb{P}_{\pinit}(\Phi)$ for all $\Phi\in\mathscr{F}_t$.

 We assume that the pair   $(X,\mathbb{P}_{\pinit})$  forms a Markov jump process with  an  initial distribution
$\pinit(u)$ and 
rates $k_{u\rightarrow v}\geq 0$ (with $u,v\in\mathcal{X}$)  that are constant in time $t$.  A Markov jump process can be represented by a  random walker that moves on the graph $G=(\mathcal{X}, \mathcal{E})$, defined by  the vertex set $\mathcal{X}$ and  the set of edges 
\begin{equation}
\mathcal{E} = \left\{(u,v)\in \mathcal{X}^2: k_{u\rightarrow v}>0\right\}. \label{eq:setEdges}
\end{equation}
The probability distribution $p_{X(t)}(u) = p(u;t)$ of $X(t)$, denoting the probability that the random walker is located at time $t$ at $X(t)=u$,  solves the differential equation  \cite{bremaud2013markov}
\begin{equation}
\partial_t p(u;t) = \sum_{v\in\mathcal{X};v\neq u}p(v;t)k_{v\rightarrow u} -  p(u;t)\sum_{v\in\mathcal{X};v\neq u}k_{u\rightarrow v}, \quad \forall u\in \mathcal{X}, \label{eq:master}
\end{equation}
with initial condition $p(u;0) = \pinit(u)$. 

We assume that the directed graph $(\mathcal{X},\mathcal{E})$ of permissible transitions on which $(X,\mathbb{P}_{\pinit})$ is defined is strongly connected, so that the stationary probability mass function  $p_{\rm ss}(u)$ that solves 
\begin{equation}
p_{\rm ss}(u) = \frac{\sum_{v\in\mathcal{X};v\neq u}p_{\rm ss}(v)k_{v\rightarrow u}}{\sum_{v\in\mathcal{X};v\neq u}k_{u\rightarrow v}}
\end{equation}
is unique (such Markov jump processes are called irreducible in Ref.~\cite{bremaud2013markov}).   If $\pinit  = p_{\rm ss}$, then we say that the Markov jump process is {\it stationary}, and we write $\mathbb{P}_{p^\ast} = \mathbb{P}_{\rm ss}$.  At stationarity,  the edge currents, as defined in Eq.~(\ref{eq:edgeCurrentDef}), are given by 
\begin{equation}
\overline{j}_{x\rightarrow y} = p_{\rm ss}(x)k_{x\rightarrow y} - p_{\rm ss}(y)k_{y\rightarrow x},
\end{equation}
and the stationarity condition $\partial_t p(u;t)=0$ implies that 
\begin{equation}
\sum_{v\in\mathcal{X};v\neq u}\overline{j}_{v\rightarrow u}=0, \quad \forall u\in \mathcal{X}. \label{eq:edgeMarkovCurrent}
\end{equation}

When the {\it microscopic affinities} 
\begin{equation}
    a_{u\rightarrow v} :=   \ln \frac{p_{\rm ss}(u)k_{u\rightarrow v}}{p_{\rm ss}(v)k_{v\rightarrow u}} \label{eq:a}
\end{equation} 
are equal to zero, i.e., 
\begin{equation}
      a_{u\rightarrow v} = 0, \quad \forall u,v\in \mathcal{E},
\end{equation}
then the Markov jump process  $(X,\mathbb{P}_{\rm ss})$ obeys {\it detailed balance}.  For Markov jump processes that obey detailed balance,   all edge currents are stalled, i.e., $\overline{j}_{u\rightarrow v} = 0$ for all $u,v\in\mathcal{X}$,  and  we say that the stationary state is an equilibrium state, which we denote   by $p_{\rm ss}(u) = p_{\rm eq}(u)$.  
  
We can also represent a Markov jump process in terms of its  trajectories $X^t_0 = \left\{X(s):s\in[0,t]\right\}$.    For a Markov jump process, the trajectory $X^t_0$ is uniquely determined by the sequence $(X_0,X_1,\ldots X_{N(t)-1})$ of $N(t)$ states that $X(t)$ occupies in the interval $[0,t]$, and the times $T_i$ when $X(t)$ changed its state from $X_{i-1}$ to $X_i$.   Indeed, it holds that 
\begin{equation}
X(s) =  X_{i}, \quad    \forall s\in [T_{i},T_{i+1}) .
\end{equation}

We denote averages  of random variables over the measure $\mathbb{P}_{\pinit}$ by $\langle \cdot \rangle_{\mathbb{P}_{\pinit}}$.   If $\pinit = p_{\rm ss}$, then we also use $\langle \cdot \rangle_{\mathbb{P}_{\rm ss}} = \langle \cdot\rangle_{\rm ss}$.

\subsection{Radon-Nikodym derivative processes}\label{sec:RN}

  Let $\mathbb{Q}_{\qinit}$ be a second probability measure defined on $(\Omega,\mathscr{F})$ for which it holds that  $(X,\mathbb{Q}_{\qinit})$ is a Markov jump process.   We denote its initial distribution by $\qinit$  and the corresponding transition rates by 
  $\ell_{u\rightarrow v}\geq 0$.

  We assume that $\mathbb{Q}_{\qinit}$ is locally, absolutely continuous with respect to $\mathbb{P}_{\pinit}$, i.e.,  
  \begin{equation}
  \mathbb{P}_{\pinit}[\Phi]=0 \Rightarrow \mathbb{Q}_{\qinit}[\Phi]=0 
    \end{equation}
for all $\Phi\in \mathscr{F}_t$ and $t\in\mathbb{R}^+$.  Locally refers here to the fact that the two measures are absolutely continuous on the sub-$\sigma$ algebras $\mathscr{F}_t$  for all {\it finite} $t$, but not necessarily on $\mathscr{F}$.    For Markov jump processes, local absolute continuity implies that that 
\begin{equation}
k_{v\rightarrow u} = 0 \Rightarrow \ell_{u\rightarrow v} = 0     \label{eq:cond1Loc}
\end{equation}
and 
\begin{equation}
p^\ast(u) =0    \Rightarrow q^\ast(u) = 0  \label{eq:cond2Loc}
\end{equation}
for all $u,v\in \mathcal{X}$.

  Since $\mathbb{Q}_{\qinit}$ is locally, absolutely continuous with respect of $\mathbb{P}_{\pinit}$, there exists a process $R(t)$, which is called the Radon-Nikodym derivative process of $\mathbb{Q}_{\qinit}$ with respect to $\mathbb{P}_{\pinit}$ \cite{liptser1977statistics}, such that
\begin{equation}
\langle f(X^t_0)\rangle_{\mathbb{Q}_{\pinit}} = \langle f(X^t_0) R(t)\rangle_{\mathbb{P}_{\pinit}}  \label{eq:Radonx} 
\end{equation}
for   measurable functions $f$ defined on the  set of trajectories $X^t_0$.    For Markov jump processes, the Radon-Nikodym derivative process takes the  form 
\begin{eqnarray}
R(t) = \frac{{\rm d}\mathbb{Q}_{\qinit}[X^t_0]}{{\rm d}\mathbb{P}_{\pinit}[X^t_0]}=  \frac{\qinit(X(0))}{\pinit(X(0))} \exp\left(\int^{t}_0 dt' \left[r_p(X(t'))-r_q(X(t'))\right] + \sum^{N(t)-1}_{i=1} \ln \frac{\ell_{X_{i-1}\rightarrow X_{i}}}{k_{X_{i-1}\rightarrow X_{i}}} \right), \label{eq:Radon} 
\end{eqnarray}
where 
\begin{eqnarray}
r_p(u) = \sum_{v\in \mathcal{X};v\neq u} k_{u\rightarrow v} \quad {\rm and} \quad r_q(u) = \sum_{v\in\mathcal{X};v\neq u} \ell_{u\rightarrow v} \label{eq:exitrates}
\end{eqnarray}  
are the escape rates out of the state $u\in\mathcal{X}$ corresponding to the measures $\mathbb{P}_{p^\ast}$ and $\mathbb{Q}_{q^\ast}$, respectively.

If the two measures have the same escape rates, i.e., 
\begin{equation}
r_p(u)=r_q(u), \quad \forall u\in\mathcal{X} \label{eq:exitId}
\end{equation}
 then we obtain the simpler expression 
\begin{eqnarray}
 \frac{{\rm d}\mathbb{Q}_{\pinit}[X^t_0]}{{\rm d}\mathbb{P}_{\pinit}[X^t_0]}= \frac{\qinit(X(0))}{\pinit(X(0))}   \exp\left( \sum_{(u, v)\in \mathcal{E}} N_{u\rightarrow v}(\omega,t) \ln \frac{\ell_{u\rightarrow v}}{k_{u\rightarrow v}} \right) ,\label{eq:Radon2}
\end{eqnarray}
where $\mathcal{E}$ is the set of pairs $(u,v)\in \mathcal{X}^2$ so that $\ell_{u\rightarrow v}> 0$, and $N_{u\rightarrow v}(t)$  is the number of times that $X$ has jumped from $u$ to $v$ in the trajectory $X^t_0$, as used before in Eq.~(\ref{eq:JEdge}).     In addition, if 
\begin{equation}
\frac{\ell_{u\rightarrow v}}{k_{u\rightarrow v}}  = \frac{k_{v\rightarrow u}}{\ell_{v\rightarrow u}}, \label{eq:rateId}
\end{equation}
then 
\begin{eqnarray}
 \frac{{\rm d}\mathbb{Q}_{\pinit}[X^t_0]}{{\rm d}\mathbb{P}_{\pinit}[X^t_0]}= \frac{\qinit(X(0))}{\pinit(X(0))}   \exp\left(\frac{1}{2} \sum_{(u, v)\in \mathcal{E}} J_{u\rightarrow v}(\omega,t) \ln \frac{\ell_{u\rightarrow v}}{k_{u\rightarrow v}} \right) ,\label{eq:Radon3}
\end{eqnarray}
which is an expression that we use later in the derivation of the main results. 

\subsection{Time-reversal in stationary Markov jump processes and  entropy production}
Time-reversal arguments  play an important role in the derivation of the main results, and it is therefore useful to revise some properties of time-reversal in Markov jump processes.    

Let $\Theta$ be the time-reversal map that mirrors trajectories with respect to the origin of time, i.e., 
\begin{equation}
X(\Theta(\omega),t) = X(-t) .
\end{equation} 
 We  define the measure
\begin{equation}
    \tilde{\mathbb{P}}_{\rm ss} = \mathbb{P}_{\rm ss}\circ\Theta
\end{equation}
of time-reversed events.   The pair  $(X, \tilde{\mathbb{P}}_{\rm ss} )$ is also a stationary Markov jump process with rates 
\begin{equation}
\tilde{k}_{u\rightarrow v} = k_{v\rightarrow u} \frac{p_{\rm ss}(v)}{p_{\rm ss}(u)}
\end{equation} 
and stationary probability mass function $\tilde{p}_{\rm ss}(u) = p_{\rm ss}(u)$.   Indeed, a direct calculation shows that~\cite{seifert2012stochastic}
\begin{eqnarray}
 \frac{{\rm d}(\mathbb{P}_{\rm ss}\circ \Theta)[X^t_0]}{{\rm d}\mathbb{P}_{\rm ss}[X^t_0]}  = e^{-S(t)}\label{eq:expder}
\end{eqnarray}  
where
\begin{equation}
    S(t) =\frac{1}{2} \sum_{(u, v)\in\mathcal{E}} a_{u\rightarrow v} J_{u\rightarrow v}(t), \label{eq:S}
\end{equation}
is the   entropy production.    Notice that the entropy production is a current of the form  Eq.~(\ref{eq:JGeneral}) with the coefficients $c_{u\rightarrow v}$ given by the microscopic affinities  $a_{u\rightarrow v}$, as defined in Eq.~(\ref{eq:a}).   

If the principle of local detailed balance  applies~\cite{seifert2012stochastic,sekimoto2010stochastic, 10.21468/SciPostPhysLectNotes.32},  which is a statistical physics implementation of local equilibrium~\cite{prigogine1978time}, then 
\begin{equation}
\dot{s} := \langle S(t)\rangle_{\rm ss}/t = \frac{1}{2} \sum_{(u, v)\in\mathcal{E}} a_{u\rightarrow v} \overline{j}_{u\rightarrow v} \label{eq:sdotDef}
\end{equation}
is the entropy production  of the second law of thermodynamics, denoting the rate of bits, measured in the natural unit of information (nat), produced on average in the environment.   In  case the environment consists of one or more thermal reservoirs, then $\dot{s}$ is  directly related to the heat dissipated to the environment.

\subsection{Martingales}\label{sec:martingales}

Radon-Nikodym derivative processes  of the form  (\ref{eq:Radon}) are martingales   with respect to the probability measure $\mathbb{P}_{\pinit}$~\cite{liptser1977statistics}.   

We use $\mathbb{P}$ to denote a generic probability measure, and not necessarily the measure $\mathbb{P}_{\pinit}$ of a Markov jump process. 
A process $M(t)$ is a $\mathbb{P}$-martingale when the following conditions hold: (i) $M(t) = M(X^t_0)$ is a functional on the trajectories of $X$; (ii)  $\langle |M(t)|\rangle_{\mathbb{P}}<\infty$; and (iii) the process is driftless, i.e., 
\begin{equation}
\langle M(t)|X^s_0\rangle_{\mathbb{P}} = M(s), \label{eq:MDrift}
\end{equation} 
for all $s\in [0,t]$. 

Martingales are  useful for studying properties of processes at random times.      This is due to  Doob's optional stopping theorem, which we briefly revisit here.    Let $T$ be a stopping time of the process $X$.   This means that $T\in[0,\infty]$ is a random time that is uniquely determined by the process $X$ and obeys causality, i.e., the stopping criterion that determines the stopping time $T$ is independent of the part of the trajectory of $X$ that takes place after the stopping time $T$.    Doob's optional stopping theorem states that when  the stopping time $T$ is finite with probability  one, and there exists a constant $c\in \mathbb{R}^+$ such that $|M(t)|<c$  for all $t\leq T$, then (see, amongst others, ~Theorem 3.6 in \cite{liptser1977statistics},  Corollary 2 in \cite{neri2019integral}, or Theorem 3.3 in \cite{bremaud2013markov})
\begin{equation}
\langle M(T)|X(0)\rangle_{\mathbb{P}} = M(0).  \label{eq:DoobStop}
\end{equation}

 Radon-Nikodym derivative process are examples of martingales, as follows readily from their definition,  see Appendix~\ref{App:A}.  Consequently, according to Eq.~(\ref{eq:expder}), the exponentiated negative entropy production is a martingale \cite{chetrite2011two, neri2017statistics,  neri2019integral, ge2021martingale}.  The martingality of $e^{-S}$ is an interesting finding for physics as it can be used to constrain the fluctuations of $S(t)$.  
For example, using the martingale property of $e^{-S}$ together with Doob's optional stopping theorem, Refs.~\cite{chetrite2011two, neri2017statistics,  neri2019integral}  derive universal laws for entropy production at stopping times $T$, including the infimum law Eq.~(\ref{eq:1}).   In the present paper, we use martingales to   determine the statistics of infima of edge currents, extending the applicability of martingales to currents that are not the entropy production.

\section{Martingales associated with edge currents}\label{sec:III}
Following  Refs.~\cite{polettini2017effective, polettini2019effective},  we identify a martingale process $M_{x\rightarrow y}$  associated with the edge current $J_{x\rightarrow y}$, which exists whenever the Markov  jump process obtained by removing the edge $x\rightarrow y$ from the original process $(X,\mathbb{P}_{\rm ss})$ has a unique stationary probability distribution.

Consider  the Markov jump process $(X,\mathbb{Q}_{\qinit})$ with rates
\begin{eqnarray}
\ell_{u\rightarrow v} = \left\{\begin{array}{ll}  k_{u\rightarrow v}, & \mathrm{for}\, (u, v)\in \left\{(x,y),(y,x)\right\}, \\ \frac{p_{\rm ss}^{x,y}(v)}{p_{\rm ss}^{x,y}(u)} k_{v\rightarrow u}, & \mathrm{for}\, (u,v)\in \mathcal{X}^2\setminus\left\{(x,y),(y,x)\right\}, \end{array}\right.    \label{hom}
\end{eqnarray} 
where $p_{\rm ss}^{x,y}$ solves the equations 
 \begin{eqnarray}
\sum'_{v\in\mathcal{X};v\neq u}k_{u\rightarrow v} =  \sum'_{v\in\mathcal{X};v \neq u}k_{v\rightarrow u}\frac{p_{\rm ss}^{x,y}(v) }{ p_{\rm ss}^{x,y}(u)}, \label{eq:pabdef}
 \end{eqnarray}
 and  $1=\sum_{v\in\mathcal{X}}p_{\rm ss}^{x,y}(v)$; the prime on the sums in Eq.~(\ref{eq:pabdef}) means that $(v,u)\notin \left\{(x,y), (y,x)\right\}$.
 
 Equation  (\ref{eq:pabdef}) implies  that the exit rates $r_p$ and $r_q$, as defined in Eq.~(\ref{eq:exitrates}), satisfy   Eq.~(\ref{eq:exitId}), and  Eq.~(\ref{hom}) implies that Eq.~(\ref{eq:rateId}) is satisfied.  Consequently, Eq.~(\ref{eq:Radon3})  applies.     Setting $\pinit = \qinit = q_{\rm ss}$, the latter being the stationary state of the process $(X,\mathbb{Q}_{\qinit})$ (which is nota bene different from $p_{\rm ss}^{x,y}$ and $p_{\rm ss}$), Eq.~(\ref{eq:Radon3})  reads 
 \begin{eqnarray}
\frac{{\rm d}\mathbb{Q}_{\rm ss}[X^t_0]}{{\rm d}\mathbb{P}_{q_{\rm ss}}[X^t_0]} &=&   \left(\frac{ p_{\rm ss}^{x,y}(x) k_{x\rightarrow y}}{p_{\rm ss}^{x,y}(y) k_{y\rightarrow x}}\right)^{-J_{y\rightarrow x}(t)} \prod_{(u,v)\in\mathcal{E};u\neq v}\left(\frac{p_{\rm ss}^{x,y}(v) k_{v\rightarrow u}}{p_{\rm ss}^{x,y}(u) k_{u\rightarrow v}}\right)^{\frac{J_{u\rightarrow v}(t)}{2}}  \nonumber\\ 
&=&\frac{p_{\rm ss}^{x,y}(X(t))p_{\rm ss}(X(0))}{p_{\rm ss}^{x,y}(X(0))p_{\rm ss}(X(t))} \left(\frac{ p_{\rm ss}^{x,y}(x) k_{x\rightarrow y}}{p_{\rm ss}^{x,y}(y) k_{y\rightarrow x}}\right)^{-J_{y\rightarrow x}(t) }  \frac{{\rm d}\mathbb{P}_{\rm ss}[\Theta (X^t_0)]}{{\rm d}\mathbb{P}_{\rm ss}[X^t_0]},    \label{eq:intermediate1}
\end{eqnarray}
where $\mathcal{E}$ is the set of permissible transitions, see Eq.~(\ref{eq:setEdges}), and we denote $\mathbb{Q}_{q_{\rm ss}} = \mathbb{Q}_{\rm ss}$ and $\mathbb{P}_{p_{\rm ss}} = \mathbb{P}_{\rm ss}$, as before.     Equation (\ref{eq:intermediate1}) is equivalent to
 \begin{eqnarray}
\frac{{\rm d}\mathbb{Q}_{{\rm ss}}[X^t_0]}{{\rm d}\mathbb{P}_{\rm ss}[\Theta (X^t_0)]} &=&  \frac{p_{\rm ss}^{x,y}(X(t))q_{\rm ss}(X(0))}{p_{\rm ss}^{x,y}(X(0))p_{\rm ss}(X(t))} \left(\frac{ p_{\rm ss}^{x,y}(x) k_{x\rightarrow y}}{p_{\rm ss}^{x,y}(y) k_{y\rightarrow x}}\right)^{-J_{y\rightarrow x}(t)}.
\end{eqnarray}
Lastly, setting $X^t_0\rightarrow \Theta(X^t_0)$, we obtain the $\mathbb{P}_{\rm ss}$-martingale  
 \begin{eqnarray}
M_{x\rightarrow y}(t) := \frac{{\rm d}\mathbb{Q}_{\rm ss}[\Theta (X^t_0)]}{{\rm d}\mathbb{P}_{\rm ss}[X^t_0]} &=&  \frac{p_{\rm ss}^{x,y}(X(0))q_{\rm ss}(X(t))}{p_{\rm ss}^{x,y}(X(t))p_{\rm ss}(X(0))} \left(\frac{ p_{\rm ss}^{x,y}(x) k_{x\rightarrow y}}{p_{\rm ss}^{x,y}(y) k_{y\rightarrow x}}\right)^{-J_{x\rightarrow y}(t)}, \label{eq:MxyT}
\end{eqnarray}
associated with the edge current $J_{x\rightarrow y}$. 

Importantly,   $M_{x\rightarrow y}(t)$ is a martingale because it is a Radon-Nikodym derivative process, and the latter are  martingales (see Appendix~\ref{App:A}).  The fact that $M_{x\rightarrow y}(t)$ is a Radon-Nikodym derivative  can also be shown directly through the identity 
\begin{equation}
M_{x\rightarrow y} =   \frac{{\rm d}\mathbb{R}_{{\rm ss}}[X^t_0]}{{\rm d}\mathbb{P}_{\rm ss}[X^t_0]}  
\label{eq:radon2x}\end{equation}
where $(X,\mathbb{R}_{\rm ss})$ is the Markov jump process with rates 
\begin{eqnarray}
m_{u\rightarrow v} = \left\{\begin{array}{ll}  \frac{q_{\rm ss}(v)}{q_{\rm ss}(u)}k_{v\rightarrow u} & \mathrm{for}\, (u, v)\in \left\{(x,y),(y,x)\right\}, \\\frac{q_{\rm ss}(v)}{q_{\rm ss}(u)} \frac{p_{\rm ss}^{x,y}(u)}{p_{\rm ss}^{x,y}(v)} k_{u\rightarrow v}, & \mathrm{for}\, (u,v)\in \mathcal{E}\setminus\left\{(x,y),(y,x)\right\}, \end{array}\right.    \label{homx}
\end{eqnarray}  
and initial distribution $p_{X(0)}(x) = q_{\rm ss}(x)$; notice that $\mathbb{R}_{{\rm ss}} = \mathbb{Q}_{\rm ss}\circ\Theta$. 
At present, the identification of (\ref{eq:MxyT}) with (\ref{eq:radon2x}) is  an insightful  exercise to convince ourselves further that $M_{x\rightarrow y}$ is a martingale, and which for completeness we present  in  Appendix~\ref{app:BBis}.

Identifying in Eq.~(\ref{eq:MxyT}) the effective microscopic affinity 
\begin{equation}
  a^\ast_{x\rightarrow y} := \ln \frac{ p_{\rm ss}^{x,y}(x) k_{x\rightarrow y}}{p_{\rm ss}^{x,y}(y) k_{y\rightarrow x}}, \label{eq:effA}
\end{equation}
as introduced in Ref.~\cite{polettini2017effective},
we obtain for the $\mathbb{P}_{{\rm ss}}$-martingale  $M_{x\rightarrow y}(t)$ the expression 
 \begin{eqnarray}
M_{x\rightarrow y}(t) &=&  \frac{p_{\rm ss}^{x,y}(X(0))q_{\rm ss}(X(t))}{p_{\rm ss}^{x,y}(X(t))p_{\rm ss}(X(0))} e^{- a^\ast_{x\rightarrow y}J_{x\rightarrow y}(t)} .  \label{eq:Mxy}
\end{eqnarray}
Notice that the effective microscopic affinity $a^\ast_{x\rightarrow y}$, defined in Eq.~(\ref{eq:effA}), has a similar form as  the  microscopic affinity, defined in Eq.~(\ref{eq:a}), with the difference that      it considers the stationary probability mass function $p_{\rm ss}^{x,y}$  in the modified process for which the  transitions  from $x$ to $y$, and vice versa, have been removed.   Although, in general, microscopic affinities are not simply related to the "true", macroscopic affinities, for   unicyclic systems it holds that the effective affinity equals the  macroscopic affinity,  while the microscopic affinity captures  in this case a  possibly small portion of the macroscopic affinity (see also Appendix~\ref{App:Cb}).

The effective affinity $a^\ast_{x\rightarrow y}$  has a  kinematic meaning.  Indeed, as we show  in Appendix~\ref{App:B},  the martingality of $M_{x\rightarrow y}(t)$ implies that 
\begin{equation}
    a^\ast_{x\rightarrow y}\langle J_{x\rightarrow y}(t)\rangle_{\rm ss} \geq 0, \label{eq:astxyPx}
\end{equation}
where the equality   holds when $a^\ast_{x\rightarrow y} = \langle J_{x\rightarrow y}(t)\rangle_{\rm ss} = 0$.

In the next section, we use the martingale $M_{x\rightarrow y}$ to determine the  statistics  of infima of $J_{x\rightarrow y}$.

\section{Statistics of   infima of edge currents} \label{sec:IV}
We determine the probability mass function of $J_{x\rightarrow y}^{\rm inf}$ for currents with   $\langle J_{x\rightarrow y}(t)\rangle_{\rm ss}>0$.  Note that in this case, according to Eq.~(\ref{eq:astxyPx}),  also $a^\ast_{x\rightarrow y}>0$.

In Secs.~\ref{sec:x0} and \ref{sec:gen}, we derive the probability mass functions of $J_{x\rightarrow y}^{\rm inf}$ for an initial state $X(0)$ that equals the source node $x$ of the edge $x\rightarrow y$ and for general initial conditions,  respectively.    In Secs.~\ref{sec:lim1} and \ref{sec:lim2}, we discuss two interesting limiting cases, namely, the case of stalled currents and the case of processes for which $(X,\mathbb{P}^{x,y}_{\rm ss})$ obeys detailed balance, respectively.

\subsection{Infimum law when the initial state equals  the source node of the edge}\label{sec:x0}
First, we determine the statistics of $J_{x\rightarrow y}^{\rm inf}$ when   the initial condition $X(0)=x$.  In this case, the derivations simplify.  

 Consider  the stopping problem  
\begin{equation}
T := {\rm inf}\left\{t\geq 0: J_{x\rightarrow y}(t) \notin (-\ell_-,\ell_+)\right\},\quad {\rm with} \quad \ell_-,\ell_+\in\mathbb{N}, \label{eq:stopping}\end{equation}
and where we use the convention that $T=\infty$ if $J_{x\rightarrow y}(t)\in (-\ell_-,\ell_+)$ for all times $t\geq0$.   Notice that since  the edge current $J_{x\rightarrow y}(t)$ is an integer-valued stochastic process, it  takes the values
\begin{equation}
J_{x\rightarrow y}(T) \in \left\{-\ell_-,\ell_+\right\}   \label{eq:JEdgeFP}
\end{equation} 
at the stopping time $T$.  
For threshold values $\ell_+\neq 0$, we have
\begin{equation}
X(T)=  \left\{\begin{array}{ccc} x, &{\rm if}& J_{x\rightarrow y}(T)=-\ell_-, \\ y, &{\rm if}& J_{x\rightarrow y}(T)= \ell_+.\end{array} \right. \label{eq:XT}
\end{equation}  
On the other hand, if $\ell_+=0$, then $T=0$ and $X(T)=X(0)=x$.      

Applying Eq.~(\ref{eq:DoobStop}) from Doob's optional stopping theorem  to the martingale $M_{x\rightarrow y}$ with initial condition $X(0) = x$, we obtain 
\begin{equation}
\frac{p_{\rm ss}^{x,y}(x)}{p_{\rm ss}(x)}\Big\langle  \frac{q_{\rm ss}(X(T))}{p_{\rm ss}^{x,y}(X(T))} e^{-a^\ast_{x\rightarrow y}J_{x\rightarrow y}(T)} \Big| X(0)=x\Big\rangle_{\rm ss}  = \frac{ q_{\rm ss}(x)}{p_{\rm ss}(x)}  , \label{eq:previous}
\end{equation}
where we have used that $M_{x\rightarrow y}(0) =   q_{\rm ss}(X(0))/p_{\rm ss}(X(0))$.   For the present setup,
\begin{equation}
\mathbb{P}_{\rm ss}\left(T<\infty|X(0)=x\right)=1,
\end{equation}
and therefore for $\ell_+\neq 0$ Eq.~(\ref{eq:previous}) reads
\begin{eqnarray}
\lefteqn{p_-\frac{ q_{\rm ss}(x)}{p_{\rm ss}(x)} \Big\langle   e^{-a^\ast_{x\rightarrow y}J_{x\rightarrow y}(T)}   | J(T)= -\ell_-,X(0)=x\Big\rangle_{\rm ss} }   && 
\nonumber \\ 
&& 
+ p_+  \frac{p_{\rm ss}^{x,y}(x)q_{\rm ss}(y)}{p_{\rm ss}^{x,y}(y)p_{\rm ss}(x)} \Big\langle    e^{-a^\ast_{x\rightarrow y}J_{x\rightarrow y}(T)}   | J(T)= \ell_+,X(0)=x\Big\rangle_{\rm ss}  =  \frac{ q_{\rm ss}(x)}{p_{\rm ss}(x)} \label{eq:next}  ,
\end{eqnarray}
where $p_+$ and $p_-$ are the so-called splitting probabilities given by
\begin{equation}
p_+  =    \mathbb{P}_{\rm ss}\left(J_{x\rightarrow y}(T)= \ell_+ |X(0)=x\right)
\end{equation}
and 
\begin{equation}
p_- =    \mathbb{P}_{\rm ss}\left(J_{x\rightarrow y}(T)= -\ell_- |X(0)=x\right). \label{eq:pm}
\end{equation}
In the limit $\ell_+\gg 1$, the second term on the left-hand side of the Eq.~(\ref{eq:next}) converges to zero, as by assumption $\langle J_{x\rightarrow y}(t)\rangle_{\rm ss}>0$ and thus also $a^\ast_{x\rightarrow y}>0$, and therefore 
\begin{equation}
p_-  =  \Big\langle   e^{-a^\ast_{x\rightarrow y}J_{x\rightarrow y}(T)}   | J(T) = -\ell_-,X(0)=x\Big\rangle^{-1}_{\rm ss} = e^{-\ell_-a^\ast_{x\rightarrow y}},  \label{eq:pMResult} 
\end{equation}
with $\ell_-\in\mathbb{N}\cup \left\{0\right\}$.  Since in  the limit of $\ell_+\rightarrow \infty$ it holds that
\begin{equation}
    p_-  = \mathbb{P}_{\rm ss}(J_{x\rightarrow y}^{\rm inf} \leq   -\ell_-|X(0)=x), \label{eq:distripM}
\end{equation}
 Eq.~(\ref{eq:pMResult}) is the cumulative distribution of $J_{x\rightarrow y}^{\rm inf}$.   Consequently,  its probability mass function reads 
\begin{equation}
p_{J_{x\rightarrow y}^{\rm inf}}(-\ell|X(0)=x) =   e^{-\ell a^\ast_{x\rightarrow y}} (1-e^{-a^\ast_{x\rightarrow y}}), \quad \forall \ell \in \mathbb{N}\cup \left\{0\right\},\label{eq:distriExpEsc}
\end{equation}
with mean value
 \begin{equation}
\langle J_{x\rightarrow y}^{\rm inf}|X(0)=x\rangle_{\rm ss} = - \frac{e^{-a^\ast_{x\rightarrow y}}}{1-e^{-a^\ast_{x\rightarrow y}}}. \label{eq:Infxxx}
\end{equation}  

When $\langle J_{x\rightarrow y}\rangle_{\rm ss}<0$,  we obtain the analogous result 
\begin{equation}
p_{J_{x\rightarrow y}^{\rm sup}}(\ell|X(0)=x) =  e^{\ell a^\ast_{x\rightarrow y}} \left(1-e^{ a^\ast_{x\rightarrow y}} \right),  \quad  \forall  \ell \in \mathbb{N}\cup \left\{0\right\},\label{eq:distriExp2}
\end{equation}
with the
mean value 
\begin{equation}
\langle J_{x\rightarrow y}^{\rm sup}|X(0)=x \rangle_{\rm ss} =  \frac{e^{a^\ast_{x\rightarrow y}}}{1-e^{a^\ast_{x\rightarrow y}}}. \label{eq:Infx2}
\end{equation}

\subsection{Infimum law for general initial conditions} \label{sec:gen} 
We follow a derivation similar to the one presented in the previous section, but now for general initial conditions  $X(0)=x_0$. 

Applying  Eq.~(\ref{eq:DoobStop}) from Doob's optional stopping theorem to the martingale $M_{x\rightarrow y}$, given by Eq.~(\ref{eq:Mxy}), we obtain 
\begin{equation}
\frac{p_{\rm ss}^{x,y}(x_0)}{p_{\rm ss}(x_0)}\Big\langle  \frac{q_{\rm ss}(X(T))}{p_{\rm ss}^{x,y}(X(T))} e^{-a^\ast_{x\rightarrow y}J_{x\rightarrow y}(T)} |X(0)=x_0\Big\rangle_{\rm ss} = \frac{ q_{\rm ss}(x_0)}{p_{\rm ss}(x_0)}  . \label{eq:previousx}
\end{equation}
  Following similar steps as those leading to Eq.~(\ref{eq:pMResult}), we obtain  in the limit of $\ell_+\gg 1$,
\begin{equation}
\mathbb{P}_{\rm ss}(J_{x\rightarrow y}^{\rm inf} \leq   -\ell_-|X(0)=x_0)   = \frac{q_{\rm ss}(x_0)}{q_{\rm ss}(x)}\frac{p_{\rm ss}^{x,y}(x)}{p_{\rm ss}^{x,y}(x_0)} e^{-a^\ast_{x\rightarrow y}\ell_-}, \quad  \forall  \ell_-\in\mathbb{N}, \label{eq:previousxx}
\end{equation}
and 
\begin{equation}
    \mathbb{P}_{\rm ss}(J_{x\rightarrow y}^{\rm inf} \leq  0|X(0)=x_0)  =1.\label{eq:previousxxa}
\end{equation} 
Notice that for general initial conditions,   Eq.~(\ref{eq:XT}) holds for values $\ell_-\neq0$ and $\ell_+\neq 0$, but not when either of the two thresholds is zero, and therefore $\ell_- \neq 0$ in Eq.~(\ref{eq:previousxx}).   From  Eqs.~(\ref{eq:previousxx}) and (\ref{eq:previousxxa}) follows that
  the probability mass function of $J_{x\rightarrow y}^{\rm inf}$, which for $\ell\in\mathbb{N}$ is given by 
\begin{equation}
p_{J_{x\rightarrow y}^{\rm inf}}(-\ell|X(0)=x_0) = \frac{q_{\rm ss}(x_0)}{q_{\rm ss}(x)}\frac{p_{\rm ss}^{x,y}(x)}{p_{\rm ss}^{x,y}(x_0)} e^{- a^\ast_{x\rightarrow y}\ell }\left(1- e^{-a^\ast_{x\rightarrow y}}\right), \label{eq:cumula1}
\end{equation}
and for $\ell=0$ it is 
\begin{equation}
p_{J_{x\rightarrow y}^{\rm inf}}(0|X(0)=x_0) = 1-\frac{q_{\rm ss}(x_0)}{q_{\rm ss}(x)}\frac{p_{\rm ss}^{x,y}(x)}{p_{\rm ss}^{x,y}(x_0)}  e^{-a^\ast_{x\rightarrow y}}.\label{eq:cumula2}
\end{equation}
Equations~(\ref{eq:cumula1}-\ref{eq:cumula2}) readily imply Eq.~(\ref{eq:distriExp}) for the probability mass function of the infimum and Eq.~(\ref{eq:Infx}) for the average value of the infimum, where we identified 
\begin{equation}
    p_{\rm esc}(x_0) = p_{J_{x\rightarrow y}^{\rm inf}}(0|X(0)=x_0). \label{eq:pescFinal}
\end{equation}
For $x_0=x$, $p_{\rm esc}$ is given by Eq.~(\ref{eq:simpleCase}). 

\subsection{Stalled currents}\label{sec:lim1}
We say that a current is stalled when  
\begin{equation}
\langle J_{x\rightarrow y}(t)\rangle_{\rm ss}= 0.
\end{equation}    
Notice that stalled currents may  exist in nonequilibrium stationary states.

For nearly stalled currents  $a^\ast_{x\rightarrow y} \approx 0$ (as shown in Appendix~\ref{App:B} and in  Refs.~\cite{polettini2017effective,polettini2019effective}), $p_{\rm esc}(x_0)\approx 0$, and $\overline{j}_{x\rightarrow y}\approx 0$.  Consequently, in this  limiting case the Eq.~(\ref{eq:distriExp}) becomes the exponential distribution
\begin{equation}
p_{J_{x\rightarrow y}^{\rm inf}}(-\ell|X(0)=x_0) = a^\ast_{x\rightarrow y}  e^{-\ell a^\ast_{x\rightarrow y}},  \quad \ell\in \mathbb{R}^+, \label{eq:distriExpx}
\end{equation}
with mean 
\begin{equation}
\langle J_{x\rightarrow y}^{\rm inf}|X(0)=x_0 \rangle_{\rm ss} = - \frac{1}{a^\ast_{x\rightarrow y}}.  \label{eq:Infxx}
\end{equation} 
Hence, the mean of the infimum diverges in the vicinity of a stalling point.

Equation~(\ref{eq:distriExpx}) implies that the statistics of infima of edge currents $J_{x\rightarrow y}$   in the vicinity of   nonequilibrium stalled states are identical to those  in the vicinity of equilibrium  states.    Therefore,  a marginal observer that only observes the edge current  $J_{x\rightarrow y}$ from the measurements of the extreme values of $J_{x\rightarrow y}$ cannot make a distinction between a nonequilibrium stalled state and an equilibrium state.

\subsection{Markov processes driven out of equilibrium by a single edge}\label{sec:lim2}
Although Ref.~\cite{companion} derives an explicit expression for the  effective rates $a^\ast_{x\rightarrow y}$  in terms of the minors of the rate matrix, this leads, in general, to a long expression without clear physical interpretation.    Here we consider a limiting case for which $a^\ast_{x\rightarrow y}$ admits a simple, explicit expression.   

We consider Markov processes $(X,\mathbb{P}_{\rm ss})$ for which the process $(X,\mathbb{P}^{x,y}_{\rm ss})$ satisfies detailed balance, where $\mathbb{P}^{x,y}_{\rm ss}$ is the measure of the stationary Markov  jump process obtained by removing the edges $x\rightarrow y$ and $y\rightarrow x$ from the original process.   Hence, in this case the stationary distribution $p^{x,y}_{\rm ss} = p_{\rm eq}$ is an  equilibrium distribution.       

Consequently, the effective affinity $a^\ast_{x\rightarrow y}$  takes  the form 
\begin{equation}
    a^\ast_{x\rightarrow y} = \ln \frac{p_{\rm eq}(x)}{p_{\rm eq}(y)} +   \ln \frac{k_{x\rightarrow y}}{k_{y\rightarrow x}} .   \label{eq:alimiting}
    \end{equation}
Parameterising 
\begin{equation}
k_{x\rightarrow y} =    \omega_{x,y} \:  p_{\rm eq}(y) \: e^{\frac{f_{x\rightarrow y}}{2\mathsf{T}_{\rm env}}},  
\end{equation} 
where $\omega_{x,y}=\omega_{y,x}$ is a symmetric kinetic parameter and $f_{x\rightarrow y} = -f_{x\rightarrow y}$ a  thermodynamic force, we obtain 
\begin{equation}
    a^\ast_{x\rightarrow y} = \frac{f_{x\rightarrow y}}{\mathsf{T}_{\rm env}}.   \label{eq:aEffT}
\end{equation}
Hence, in the present  case, the microscopic affinity is directly related to the  thermodynamic force $f_{x\rightarrow y}$.

\section{Illustration of infimum laws for   two-headed molecular motors} \label{sec:kinesin}

We use a model for two-headed molecular motors to illustrate the  implications of the infimum laws for edge currents on  the dynamics of a physical process.  We first  introduce  in Sec.~\ref{sec:VIA} a Markov jump process for molecular motor dynamics, and subsequently in Sec.~\ref{VI:C},  we use this model to study the  extreme values  in the position of  molecular motors.

 \begin{figure}[t!]\centering
 \includegraphics[width=0.5\textwidth]{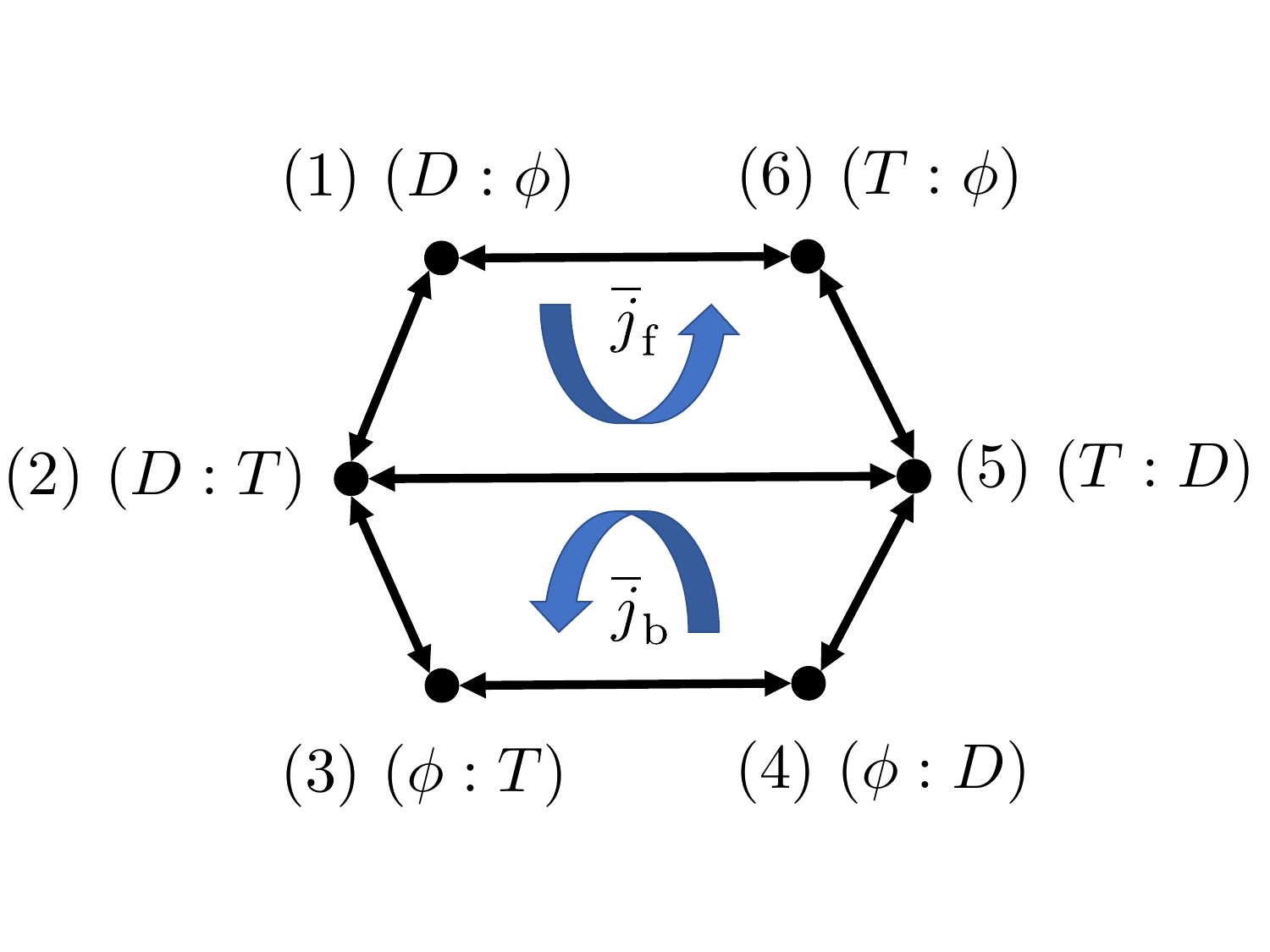}
\caption{Six-state model for molecular motors with two motor heads [figure taken from \cite{neri2}].} \label{fig:examples}
\end{figure}  

\subsection{Model for two-headed molecular motors}\label{sec:VIA}
Consider a molecular motor bound to a one-dimensional substrate that walks with  discrete steps of size $\pm \delta$.   The molecular motor is driven out of equilibrium by two thermodynamic forces, namely, an input of free energy $\Delta \mu$  -- due to the hydrolysis of adenosine triphosphate (ATP) into adenosine diphosphate (ADP) and an inorganic phosphate (P) ---  and a mechanical force $f_{\rm mech}$.    The  average entropy production rate is 
\begin{equation}
\dot{s} = \frac{\langle S(t)\rangle_{\rm ss}}{t} = \frac{ \Delta \mu}{\mathsf{T}_{\rm env}}\overline{j}_{\rm fuel} -\frac{f_{\rm mech}\delta}{\mathsf{T}_{\rm env}}  \overline{j}_{\rm pos} ,  \label{eq:sdotx}
\end{equation}
where $\mathsf{T}_{\rm env}$ is the temperature of the environment, $\overline{j}_{\rm fuel}$ is the average rate of the reaction ATP$\rightarrow$ ADP+P minus the rate of the reverse reaction ADP+P$\rightarrow$ ATP, and $\overline{j}_{\rm pos}$ is the average rate at which the motor moves forwards minus the rate at which the motor moves backwards.  The minus sign in front of $f_{\rm mech}$ indicates that the mechanical force pushes the motor in the negative direction.     The free energy associated with the hydrolysis reaction is given by 
\begin{equation}
 \Delta \mu = \mathsf{T}_{\rm env}   \ln  \left(K_{\rm eq} \frac{[{\rm ATP}]}{[{\rm ADP}][{\rm P}] }\right), \label{eq:deltamu}
\end{equation}
where $K_{\rm eq}$ is the equilibrium constant of the hydrolysis interaction,  and $[{\rm ATP}]$, $[\rm ADP]$ and $[{\rm P}]$ are, respectively, the concentrations of  ${\rm ATP}$, ${\rm ADP}$ and ${\rm P}$ in the surrounding medium.

In what follows,  we consider the six-state model for two-headed molecular motors as introduced in  Ref.~\cite{Liepelt}, which is a  Markov jump process that describes the basic features of the molecular motor's thermodynamics as described by Eq.~(\ref{eq:sdotx}).    In this model,   the position of the molecular motor along the biofilament is proportional to an edge current, and hence the theory for extreme values of Sec.~\ref{sec:IV} applies.

The six states   represent the different chemical states of the rear and front motor heads, both of which can be in an  ATP-bound state (T),  ADP-bound state (ADP) and  nucleotide-free state ($\phi$).   Since the motor heads move out of phase, the  three states $(\phi:\phi)$, $(D:D)$, and $(T:T)$ are excluded, and the process takes six possible states, 
\begin{equation}
X(t) \in \left\{ ({\rm D}:\phi), ({\rm T}:\phi), ({\rm T}:D), (\phi:{\rm D}), (\phi:{\rm T}), ({\rm D}:{\rm T})\right\}.
\end{equation}
For convenience, we also label states by 1 to 6, as indicated in Fig.~\ref{fig:examples}. 
   The pairs of states $(D:\phi)$ and $(\phi:D)$ --- but also $(D:T)$ and $(T:D)$, or  $(T:\phi)$ and $(\phi:T)$ --- are not identical, as in the state $(D:\phi)$ the rear motor head is bound to ADP and the front motor head is in the nucleotide free state, while in $(\phi:D)$ it is the other way around.       The asymmetry in the configurations  $(D:\phi)$  and $(\phi:D)$  is due to an asymmetry in the periodic, electric potential of the  one-dimensional substrate to which the motor is bound.

The dynamics of $X(t)$ is governed by a Markov jump process  with the   nonzero transition rates indicated by arrows in Fig.~\ref{fig:examples}.   All transitions, except those between  $({\rm D}:{\rm T})$ and  $({\rm T}:{\rm D})$, are chemical transitions.         On the other hand, the transition from $({\rm D}:{\rm T})$ to $({\rm T}:{\rm D})$, and vice-versa, is a mechanical transition where the motor heads swap position.      Therefore, the position $J_{\rm pos}(t)$ of the motor is the edge current
 \begin{equation}
 J_{\rm pos}(t) := J_{({\rm T}:{\rm D})\rightarrow ({\rm D}:{\rm T})}(t) = J_{2\rightarrow 5}(t), \label{eq:edgePos}
 \end{equation}
 where consistently with the setup of Sec.~\ref{sec:II},  we have set $ J_{\rm pos}(0)=0$.

Following Refs.~\cite{Liepelt, hwang2018energetic}, we parameterise the jump rates corresponding to the mechanical  transitions as 
\begin{equation}
k_{2\rightarrow 5}(f_{\rm mech}) =  k_{2\rightarrow 5}(0) e^{-\theta\frac{f_{\rm mech} \delta}{\mathsf{T}_{\rm env}}}  \label{eq:param1}
 \end{equation}
 and 
 \begin{equation}
 k_{5\rightarrow 2}(f_{\rm mech}) =  k_{5\rightarrow 2}(0) e^{(1-\theta) \frac{f_{\rm mech} \delta}{\mathsf{T}_{\rm env}}}\label{eq:param2},
  \end{equation}  
  and the chemical transitions are parameterised as
  \begin{equation}
  k_{i\rightarrow j}(f_{\rm mech}) = \frac{2k_{i\rightarrow j}(0)}{1+e^{\chi_{ij}\frac{f_{\rm mech}\delta}{\mathsf{T}_{\rm env}}}}\label{eq:param3}
  \end{equation}
  with $\chi_{ij} = \chi_{ji}$.   Note that also the chemical transitions depend on the mechanical force $f_{\rm mech}$, as the force will deform the motor heads affecting the rate of chemical reactions.     However, since $\chi_{ij}=\chi_{ji}$, it holds that 
    \begin{equation}
     \frac{k_{i\rightarrow j}(f_{\rm mech})}{k_{j\rightarrow i}(f_{\rm mech})}  =  \frac{k_{i\rightarrow j}(0)}{k_{j\rightarrow i}(0)} \label{eq:Ratios}
  \end{equation}
  for $(i,j)\notin \left\{(2,5),(5,2)\right\}$ and  all values of $f_{\rm mech}$, and consequently $f_{\rm mech}$ provides a nonzero contribution to the microscopic affinity $a_{2\rightarrow 5}$ only.  
   
  
  The concentrations of $[{\rm ADP}]$ and $[{\rm P}]$ are assumed to be constant, and the dependence on [${\rm ATP}$] enters into the model through 
  \begin{equation}
k_{1\rightarrow 2}(0) = k^{\rm bi}_{1\rightarrow 2}[{\rm ATP}] \label{eq:atp1}
  \end{equation}
  and 
   \begin{equation}
k_{4\rightarrow 5}(0) = k^{\rm bi}_{4\rightarrow 5}[{\rm ATP}], \label{eq:atp2}
  \end{equation}   
  where the $ k^{\rm bi}_{1\rightarrow 2}$ and  $k^{\rm bi}_{4\rightarrow 5}$ are rate constants whose value depend on the properties of the motor.
Due to the equivalence of transitions in the backward and forward cycles, we set $k_{3\rightarrow 2}(0) = k_{6\rightarrow 5}(0)$, $k_{2\rightarrow 3}(0) = k_{5\rightarrow 6}(0)$, $k_{3\rightarrow 4}(0) = k_{6\rightarrow 1}(0)$, $k_{4\rightarrow 3}(0) = k_{1\rightarrow 6}(0)$, $k_{4\rightarrow 5}(0) = k_{1\rightarrow 2}(0) $, $\chi_{23}=\chi_{56}$, $\chi_{34}=\chi_{61}$, and $\chi_{45}=\chi_{12}$. 
 In addition,  following Ref.~\cite{neri2}, we use
 \begin{equation}
    k_{5\rightarrow 2}(0) = k_{2\rightarrow 5}(0) \sqrt{\frac{k_{5\rightarrow 4}(0)}{k_{2\rightarrow 1}(0)}}, \label{eq:K52}
\end{equation}
so that the six-state model satisfies detailed balance when 
\begin{equation}
f_{\rm mech}=0  \quad {\rm and} \quad  [{\rm ATP}] =  \frac{k_{1\rightarrow 6}(0)}{k_{6\rightarrow 1}(0)} \frac{k_{6\rightarrow 5}(0)}{k_{5\rightarrow 6}(0)} \frac{\sqrt{k_{5\rightarrow 4}(0)k_{2\rightarrow 1}(0)} }{k^{\rm bi}_{4\rightarrow 5}}, \label{eq:eqcond}  
\end{equation}
and accordingly  both thermodynamic forces $f_{\rm mech}=\Delta \mu=0$ at these values of the control parameters.  The stationary distribution $p_{\rm ss} = p_{\rm eq}$ at equilibrium is presented in Appendix~\ref{App:C}. 
The remaining constants $k_{i\rightarrow j}(0)$, $\chi_{ij}$, and $\theta$ can be   determined by fitting the model to   single molecule motility data, and for Kinesin-1 we report these values in  Appendix~\ref{app:C}.

As shown in Fig.~\ref{fig:examples}, the model has three  cycles, one corresponding  to a forward motion at a rate $ \overline{j}_{\rm f} $,  one corresponding to a backward motion at a rate $\overline{j}_{\rm b}$, and one for which the motor does not move but hydrolyses two ATP molecules into ADP and P at a rate $\overline{j}_0$.  The corresponding thermodynamic affinities are 
\begin{equation}
\frac{a_{\rm f}}{\mathsf{T}_{\rm env}}   =  \ln \frac{k_{1\rightarrow 2}k_{2\rightarrow 5}k_{5\rightarrow 6}k_{6\rightarrow 1}}{k_{1\rightarrow 6}k_{6\rightarrow 5}k_{5\rightarrow 2}k_{2\rightarrow 1}} =  \frac{\Delta \mu}{\mathsf{T}_{\rm env}}  - \frac{f_{\rm mech} \delta}{\mathsf{T}_{\rm env}}, 
\end{equation} 
\begin{equation}
\frac{a_{\rm b}}{\mathsf{T}_{\rm env}}   =  \ln \frac{k_{2\rightarrow 3}k_{3\rightarrow 4}k_{4\rightarrow 5}k_{5\rightarrow 2}}{k_{2\rightarrow 5}k_{5\rightarrow 4}k_{4\rightarrow 3}k_{3\rightarrow 2}} =  \frac{\Delta \mu}{\mathsf{T}_{\rm env}} + \frac{f_{\rm mech} \delta}{\mathsf{T}_{\rm env}},
\end{equation}
and 
\begin{equation}
\frac{a_{0}}{\mathsf{T}_{\rm env}}   =  \ln \frac{k_{2\rightarrow 3}k_{3\rightarrow 4}k_{4\rightarrow 5}k_{5\rightarrow 6}k_{6\rightarrow 1}k_{1\rightarrow 2}}{k_{3\rightarrow 2}k_{2\rightarrow 1}k_{1\rightarrow 6}k_{6\rightarrow 5}k_{5\rightarrow 4}k_{4\rightarrow 3}} =  \frac{2\Delta \mu}{\mathsf{T}_{\rm env}} .
\end{equation}
The  total, average rate of dissipation, as defined in Eq.~(\ref{eq:sdotDef}), is thus given by 
\begin{equation}
\dot{s} = \overline{j}_{\rm f} \frac{a_{\rm f}}{\mathsf{T}_{\rm env}}   + \overline{j}_{\rm b} \frac{a_{\rm b}}{\mathsf{T}_{\rm env}} + \frac{a_0}{\mathsf{T}_{\rm env}}\overline{j}_0 ,  
\end{equation}
which provides an alternative decomposition of the average entropy production rate from Eq.~(\ref{eq:sdotx}).

   \begin{figure}[t!]\centering
 \includegraphics[width=0.4\textwidth]{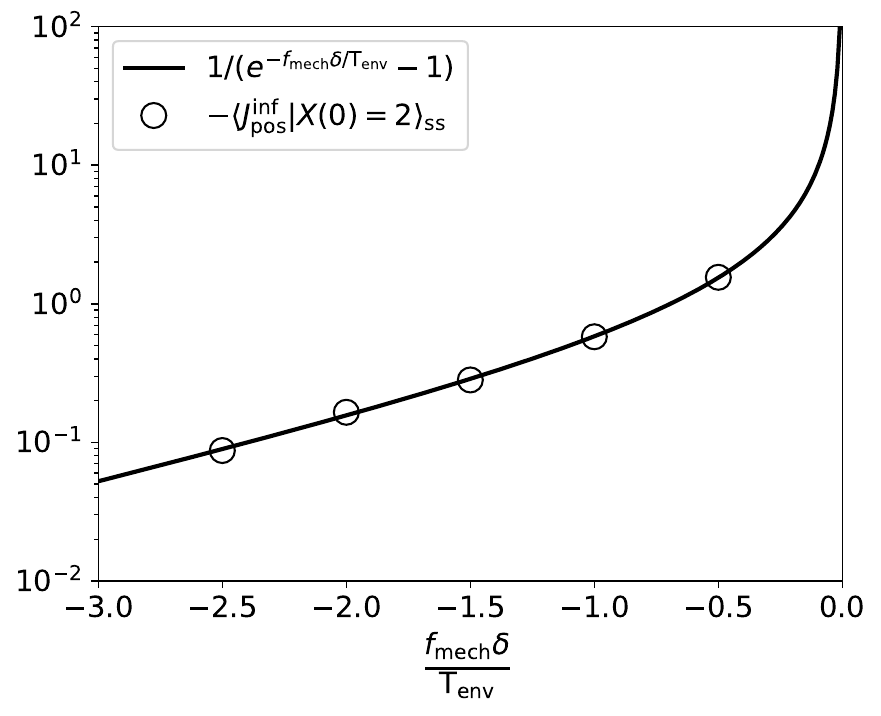}
\caption{Mean infimum for the position  $J_{\rm pos}(t) = J_{2\rightarrow 5}(t)$  of a molecular motor in the model of Sec.~\ref{sec:VIA} (also illustrated in Fig.~\ref{fig:examples}) as a function of the   mechanical force $f_{\rm mech}$ and without  chemical driving, i.e.,  $\Delta \mu=0$.   Theoretical results given by Eq.~(\ref{eq:exreme})   (line) are compared with empirical averages from numerical simulations (markers).   Simulation results are empirical averages for the most negative value of the position of the molecular motor averaged over $5e+3$ realisations of the process when the initial state  $X(0)=2$.  } \label{fig:results1}
\end{figure}   

\subsection{Infimum laws for the position of molecular motors}\label{VI:C} 

We determine the statistics of the infima  $J_{\rm pos}^{\rm inf}$ in the position $J_{\rm pos}$ of  molecular motors, as described by the six state model illustrated in Fig.~\ref{fig:examples}.    Since $J_{\rm pos}$ is the edge current corresponding to the $2\rightarrow 5$ transition, see Eq.~(\ref{eq:edgePos}), this boils down, according to Eqs.~(\ref{eq:distriExp}) and (\ref{eq:Infx}), to evaluating the effective affinity 
\begin{equation}
a^\ast_{2\rightarrow 5} = \ln \frac{p^{2,5}_{\rm ss}(2)}{p^{2,5}_{\rm ss}(5)} + \ln \frac{p_{\rm eq}(5)}{p_{\rm eq}(2)} - \frac{f_{\rm mech}\delta}{\mathsf{T}_{\rm env}} . \label{eq:aast25x}
\end{equation}
To derive (\ref{eq:aast25x}), we have used the expression (\ref{eq:effA}) for the effective affinity together with the  detailed balance condition 
\begin{equation}
\frac{k_{2\rightarrow 5}(0)}{k_{5\rightarrow 2}(0)} = \frac{p_{\rm eq}(5)}{p_{\rm eq}(2)}
\end{equation} 
satisfied by the rates $k_{2\rightarrow 5}(0)$ and $k_{5\rightarrow 2}(0)$.    The  distribution $p_{\rm eq}(x)$ is the stationary distribution at equilibrium conditions $\Delta \mu = f_{\rm mech} = 0$, and  the explicit values of $p_{\rm eq}$ are presented  in Appendix~\ref{App:C}.

\begin{figure*}[t!]\centering
  \includegraphics[width=0.43\textwidth]{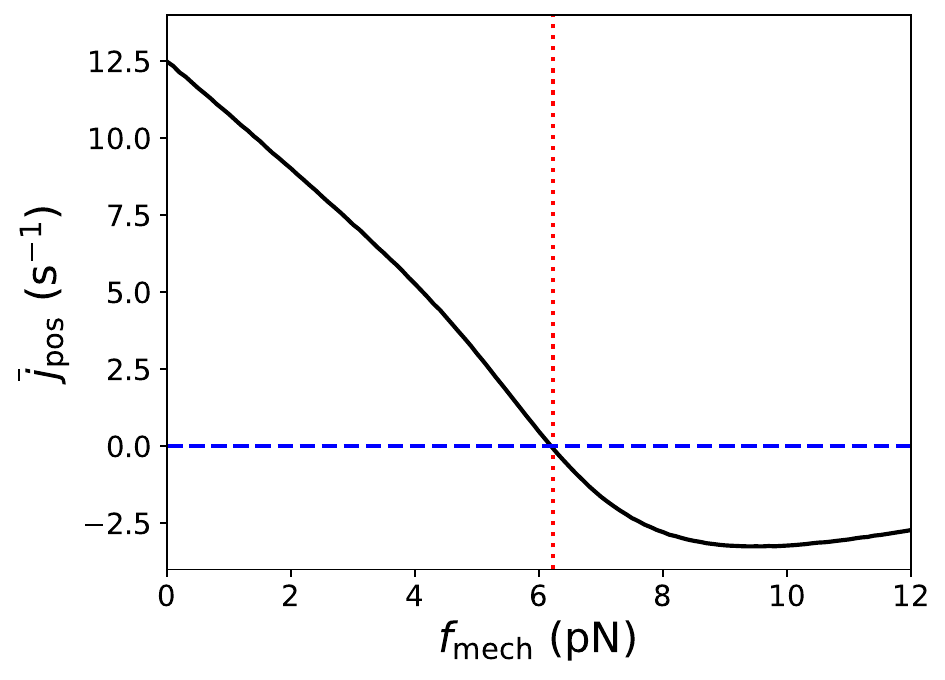}
                            \put(-30,115){\Large (a)} 
 \includegraphics[width=0.4\textwidth]{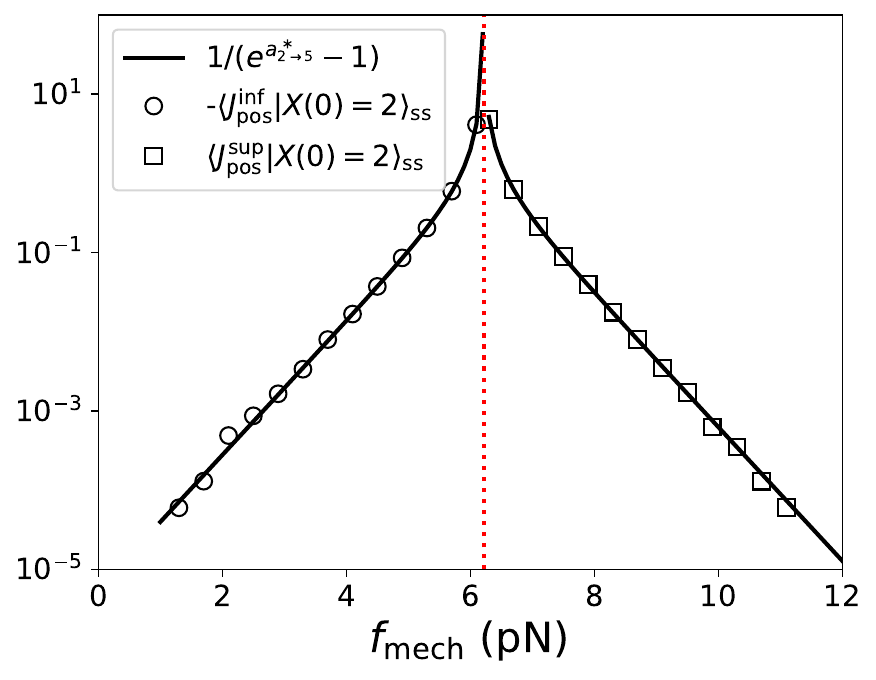}
                 \put(-30,115){\Large (b)} 
\caption{ Illustration of the infimum law on  the position $J_{\rm pos}(t) = J_{2\rightarrow 5}(t)$ of the molecular motor model  for Kinesin-1  illusrated in Fig.~\ref{fig:examples}.    Parameters used are as described in  Sec.~\ref{sec:kinesin} with  $[{\rm ATP}] = 10\mu {\rm M}$.  (a) Number of molecular motor steps per second $\overline{j}_{\rm pos}$ as a function of the mechanical force $f_{\rm mech}$;  (b) mean  infimum $\langle J_{\rm pos}^{\rm inf}|X(0)=2\rangle_{\rm ss}$ ($f_{\rm mech}<f_{\rm s}$) or supremum $\langle J_{\rm pos}^{\rm sup}|X(0)=2\rangle_{\rm ss}$  ($f_{\rm mech}>f_{\rm s}$) as a function of the mechanical force $f_{\rm mech}$, where $f_{\rm s}\approx 6.2 {\rm pN}$ is the stalling force.    Theoretical curves (lines), obtained from plotting the Eq.~(\ref{eq:Infx}),  are compared with results from  continuous-time Monte-Carlo simulations (markers); each marker is the  sample average over $10^5$ realisations of the process.  The stalling force is denoted by a vertical dotted line. } \label{fig:results}
\end{figure*}

In what follows, we determine $a^\ast_{2\rightarrow 5}$ in three cases: (i) chemical equilibrium but mechanical driving ($\Delta \mu =0, |f_{\rm mech}|>0$); (ii) mechanical equilibrium but chemical driving  ($|\Delta \mu| >0, f_{\rm mech}=0$); (iii) mechanical and chemical driving  ($|\Delta \mu| >0, |f_{\rm mech}|>0$).    In cases (i) and (ii) the stalled state is the equilibrium state, while in the latter the stalled state is, in general, a nonequilibrium state.  

\subsubsection{Chemical equilibrium and mechanical driving} 
We consider a molecular motor in chemical equilibrium  with its environment --- $[\rm ATP]$ is given by Eq.~(\ref{eq:eqcond}), such that  $\Delta \mu=0$ --- and that is driven out of equilibrium by a mechanical force~$f_{\rm mech}$.   
The limiting case of Sec.~\ref{sec:lim2} applies here, i.e.,  $p^{(2,5)}_{\rm ss} = p_{\rm eq}$, as $\chi_{ij} = \chi_{ji}$, and hence the  force $f_{\rm mech}$ does not  change the ratios of the transition rates, except for the transition from $2$ to $5$, and vice versa.    Consequently,
the  effective affinity  is determined by the thermodynamic force through
\begin{equation}
a^\ast_{2\rightarrow 5} =  -\frac{f_{\rm mech}\delta}{\mathsf{T}_{\rm env}}. \label{eq:ast}
\end{equation} 

Using Eq.~(\ref{eq:ast}) in Eqs.~(\ref{eq:Infx}) and (\ref{eq:Infx2}), we obtain explicit analytical expressions for the mean extreme values of the molecular motor position, viz.,
\begin{equation}
    \langle J_{\rm pos}^{\rm inf} |X(0)=2\rangle_{\rm ss} =  -\frac{1}{e^{-\frac{f_{\rm mech}\delta}{\mathsf{T}_{\rm env}}}-1} \quad {\rm and}  \quad  \langle J_{\rm pos}^{\rm sup}|X(0)=2 \rangle_{\rm ss} =  \frac{1}{e^{\frac{f_{\rm mech}\delta}{\mathsf{T}_{\rm env}}}-1} \label{eq:exreme}
\end{equation}
for $f_{\rm mech}<0$ and  $f_{\rm mech}>0$, respectively.   

In Fig.~\ref{fig:results1},  we plot  Eq.~(\ref{eq:exreme}) as a function of $f_{\rm mech}$ together with  results from  continuous-time Monte Carlo simulations obtained from empirical averages over  $10^4$ trajectories.    
Notice the divergence  of the mean value of the infimum when approaching the equilibrium state $f_{\rm mech}\rightarrow 0$.

\subsubsection{Mechanical equilibrium and chemical driving}
Now, we consider  the opposite case for which the motor is driven out of equilibrium by  chemical fuel, while the mechanical force equals zero. 

In this case,  the effective affinity    $a^\ast_{2\rightarrow 5}$ does not admit a simple thermodynamic interpretation in terms of the nonequilibrium forcing $\Delta \mu$.   Indeed, the  thermodynamic force $\Delta \mu$ acts on two edges, namely, $1\rightarrow 2$ and $4\rightarrow 5$, and hence the limiting case of  Sec.~\ref{sec:lim2} does not apply here.  Consequently, also    Eq.~(\ref{eq:aEffT})  that expresses the effective affinity in terms of the thermodynamic force does  not hold.

The fact that   $a^\ast_{2\rightarrow 5}$ does not admit a simple thermodynamic expression in terms of the thermodynamic force $\Delta \mu$  is  even true when $\Delta \mu \rightarrow 0$.  Indeed, in the linear response regime
\begin{eqnarray}
a^\ast_{2\rightarrow 5} =
\frac{1}{p_{\rm eq}(5)k_{5\rightarrow 2}(0)}\frac{\Delta \mu}{\mathsf{T}_{\rm env}}\left( k_{2\rightarrow 5}(0)p^{2,5}_{\mu}(2) - k_{5\rightarrow 2}(0)p^{2,5}_{\mu}(5) \right)  + O\left(\left(\frac{\Delta \mu}{\mathsf{T}_{\rm env}}\right)^2\right),
\end{eqnarray} 
which does not admit a simple interpretation  in terms of the nonequilibrium forcing $\Delta \mu$ and the Onsager coefficients of the currents in the process $(X,\mathbb{P}_{\rm ss})$.

\subsubsection{Chemical and mechanical driving}
Lastly, we discuss  extreme values  in the vicinity of   nonequilibrium, stalled states.   We set $[\rm ATP] = 10\mu M$, in which case  the motor stalls at a force  $f_{\rm mech}=f_{\rm s}\approx 6.2 {\rm pN}$, as shown in Panel (a) of Fig.~\ref{fig:results}. The average rate of dissipation $\dot{s}>0$, and hence this is a nonequilibrium stalled state for which the motor does not realise average motion despite constantly consuming chemical energy.   Nevertheless, as shown in Panel (b) of Fig.~\ref{fig:results}, the mean value of the  extreme value of $J_{\rm pos}$ diverges near $f_{\rm mech}=f_{\rm s}$, similar to the mean extreme value near equilibrium shown in Fig.~\ref{fig:results1}.    This follows from the fact that the statistics of infima, as determined by Eqs.~(\ref{eq:distriExpx}) and (\ref{eq:Infxx}), are the same for  stalled states and for equilibrium states.

\section{Estimating the average  entropy  production rate  based on  the extreme value statistics of   an edge current } \label{sec:appl}
  Given   that average entropy production rates can be estimated from the fluctuations of currents at a fixed time, see Ref.~\cite{gingrich2017inferring}, it is  natural to expect that  average  entropy production rates can  also be estimated with the extreme value statistics of currents.   In this Section, we introduce two estimators for dissipation based on extreme value statistics of currents, namely, $\hat{s}_{\rm inf}$ that applies to arbitrary currents $J$, and $\doublehat{s}_{\rm inf}$ that applies to edge currents $J_{x\rightarrow y}$. We compare the bias of the estimators $\hat{s}_{\rm inf}$ and $\doublehat{s}_{\rm inf}$   with  estimators  that have been studied previously in the literature, in particular, the thermodynamic uncertainty ratio $\hat{s}_{\rm TUR}$~\cite{barato2015thermodynamic, PhysRevE.96.020103, gingrich2017inferring} and a naive estimator $\hat{s}_{\rm KL}$ based on neglecting nonMarkovian statistics in the Kullback-Leibler divergence of the integrated edge current~\cite{gomez2008lower, roldan2010estimating, uhl2018fluctuations, martinez2019inferring, harunari2022learn}.

We start with reviewing in Sec.~\ref{sec:81}  estimators of~$\dot{s}$ that have been studied previously in the literature.  Subsequently, in Sec.~\ref{sec:82},   we discuss the two  estimators  $\hat{s}_{\rm inf}$ and $\doublehat{s}_{\rm inf}$ that are based on extreme value statistics.    Lastly, in Sec.~\ref{sec:83}, we evaluate  the quality of the different estimators of dissipation when applied to  the current $J_{\rm pos}$ of the molecular motor model defined in   Sec.~\ref{sec:VIA}.  
\subsection{Estimators of the average entropy production rate revisited}\label{sec:81}
To evaluate the quality of estimators that are based on extreme value statistics, we first review   three   well-studied estimators of dissipation, all of which are evaluated on  the trajectories  $J^t_0$ of an arbitrary  current $J$, as defined in Eq.~(\ref{eq:JGeneral}):
\begin{enumerate}
\item  The {\it Kullback-Leibler divergence} $\hat{s}_{\rm KL}$:
Let $\mathcal{J}$ be the set of jump sizes of the current $J(t)$, excluding  jumps of size zero.   Then, the Kullback-Leibler divergence of the current $J$, neglecting non-Markovian statistics, is defined by   \cite{gomez2008lower, roldan2010estimating, uhl2018fluctuations, martinez2019inferring,van2022thermodynamic,  harunari2022learn}
\begin{equation}
    \hat{s}_{\rm KL} := \sum_{j\in \mathcal{J}}\dot{n}_{j} \ln \frac{\dot{n}_j}{\dot{n}_{-j}} ,  \label{eq:KL}
\end{equation}
where $\dot{n}_j$ denotes the rate at which the current $J$ makes jumps of size $J(t)-J(t^-) = j$, and where $t^-$ denotes a time infinitesimal smaller than $t$.   The estimator $ \hat{s}_{\rm KL}$ is obtained from the  Kullback-Leibler divergence 
\begin{equation}
   \Big \langle  \ln \frac{{\rm d}\mathbb{P}_{\rm ss}[J^t_0]}{{\rm d}(\mathbb{P}_{\rm ss}\circ\Theta)[J^t_0]} \Big\rangle_{\rm ss}\label{eq:KLJ}
\end{equation}
by ignoring non-Markovian statistics in the trajectory of  $J^t_0$; notice that in Eq.~(\ref{eq:KLJ}) $\mathbb{P}_{\rm ss}[J^t_0]$  denotes the probability measure $\mathbb{P}_{\rm ss}$ constrained to the $\sigma$-algebra generated by $J^t_0$.   Since time-reversal flips the sign of the current, i.e., $J(\Theta(\omega),t) = -J(\omega,t)$, we obtain in the logarithm of Eq.~(\ref{eq:KL}) the ratio between $n_j$ and $n_{-j}$.

The Kullback-Leibler divergence lower bounds $\dot{s}$, i.e.,
\begin{equation}
    \hat{s}_{\rm KL}\leq \dot{s}.
\end{equation}  
However, when the statistics of the current $J$ contain strong non-Markovian effects and when $J$ is not proportional to the entropy production $S$, than   $\hat{s}_{\rm KL}$ provides a poor estimate of $\dot{s}$ as it does not capture the irreversibility in the non-Markovian statistics~\cite{gomez2008lower, roldan2010estimating, uhl2018fluctuations, martinez2019inferring, van2022thermodynamic, harunari2022learn}.  

\item  
 The {\it thermodynamic uncertainty ratio } $\hat{s}_{\rm TUR}$: this ratio is   defined by  \cite{barato2015thermodynamic, PhysRevE.96.020103, gingrich2017inferring}
\begin{equation}
    \hat{s}_{\rm TUR} := 2 \frac{\overline{j}^2 t}{\sigma^2_{J(t)}}, 
\end{equation}
where $\overline{j} = \langle J(t)\rangle_{\rm ss}/t$ and $\sigma^2_{J(t)} = \langle J^2(t)\rangle_{\rm ss} - \langle J(t)\rangle^2_{\rm ss}$.     For Markov jump processes the thermodynamic uncertainty ratio lower bounds $\dot{s}$, i.e.,
\begin{equation}
\hat{s}_{\rm TUR}\leq \dot{s},
\end{equation}
see Refs.~\cite{barato2015thermodynamic, gingrich2015dissipation, pietzonka2015universal, PhysRevE.96.020103}.
However,  in Markov jump processes that are governed far from thermal equilibrium,  $\hat{s}_{\rm TUR}/\dot{s}\approx 0$ \cite{neri2},  and hence  the thermodynamic uncertainty ratio captures a negligible fraction of the dissipation in this limit.  Notice that in overdamped Langevin processes $\hat{s}_{\rm TUR}/\dot{s}\approx 1$ for small $t$, as for example shown in Ref.~\cite{PhysRevLett.124.120603}.  However,  this  relies on the fact that the distribution of $S(t)$ is Gaussian  for small $t$, which does not apply to processes with jumps.   

\item  The {\it first-passage  ratio } $\hat{s}_{\rm FPR}$:  Let $T = {\rm inf}\left\{t\geq 0: J(t) \notin (-\ell_-,\ell_+)\right\}$ be the first time a current $J$ exits an open interval $(-\ell_-,\ell_+)$, and let us assume $\langle J(t)\rangle_{\rm ss} >0$. 
The fist-passage ratio of the current $J$ is defined by  \cite{roldan2015decision, neri2021universal, neri2}
\begin{equation}
    \hat{s}_{\rm FPR}(\ell_+,\ell_-) := \frac{\ell_+}{\ell_-}\frac{| \ln p_-|}{\langle T\rangle_{\rm ss}}, \label{eq:FPR}
\end{equation}
where $p_- = \mathbb{P}_{\rm ss}\left(J(T)\leq -\ell_-\right)$ is the probability that the current goes below the threshold $-\ell_-$ before exceeding the threshold $\ell_+$.   Ref.~\cite{neri2021universal} shows that in the limit of large thresholds $\ell_-$ and $\ell_+$, while keeping the ratio $\ell_-/\ell_+$ fixed,     
\begin{equation}
\hat{s}_{\rm FPR}\leq \dot{s}.  
\end{equation}
In addition, when $J$ is proportional to $S$, then  in the same limit $\hat{s}_{\rm FPR}= \dot{s}$. Although results in Ref.~\cite{neri2} indicate that in general the bias of $\hat{s}_{\rm FPR}$ is smaller than the bias in $\hat{s}_{\rm KL}$ and $\hat{s}_{\rm TUR}$, the estimator $\hat{s}_{\rm FPR}$ has the drawback that it should be evaluated at large thresholds $\ell_-$ and $\ell_+$.   

In the case of  $J  = J_{x\rightarrow y}$, we can,   using the martingale methods discussed in this paper, evaluate the bias of the estimator $\hat{s}_{\rm FPR}$ in the limit of large thresholds.  Indeed,  as shown in Ref.~\cite{neri2021universal}, 
\begin{equation}
    \langle T\rangle_{\rm ss} =\frac{\ell_+}{\overline{j}_{x\rightarrow y}} (1+o_{\ell_{\rm min}}(1)),
\end{equation}
where $o_{\ell_{\rm min}}(1)$  is the little-o notation that denotes an arbitrary function that converges to zero when both  $\ell_+$ and $\ell_-$ diverge while their ratio is kept fixed, and  $\overline{j}_{x\rightarrow y} = \langle J_{x\rightarrow y}(t)\rangle_{\rm ss}/t$.   Additionally, Eq.~(\ref{eq:distriExp}) implies that for $\overline{j}_{x\rightarrow y}>0$,
\begin{equation}
\frac{|\ln p_-|}{\ell_-} = a^\ast_{x\rightarrow y}(1+o_{\ell_{\rm min}}(1)).    \label{eq:PM}
\end{equation} 
Using Eqs.~(\ref{eq:distriExp}) and (\ref{eq:PM}) in (\ref{eq:FPR}), we obtain \begin{equation}
    \hat{s}_{\rm FPR}  =  a^\ast_{x\rightarrow y}\overline{j}_{x\rightarrow y}(1+o_{\ell_{\rm min}}(1)).  \label{eq:FPRBiasx}
\end{equation}
Hence, the average rate of dissipation estimated by a marginal observer is  the current rate $\overline{j}_{x\rightarrow y}$ times the effective affinity $a^\ast_{x\rightarrow y}$, which justifies calling     $a^\ast_{x\rightarrow y}$ an effective affinity.   It follows from Eq.~(\ref{eq:ineqS}) that 
\begin{equation}
    a^\ast_{x\rightarrow y}\overline{j}_{x\rightarrow y} \leq \dot{s},   \label{eq:ain}
\end{equation}
which has also been derived in Ref.~\cite{bisker2017hierarchical, polettini2019effective} using a different approach.

\end{enumerate}

\subsection{Estimators of dissipation based on  infimum  statistics}\label{sec:82}
We introduce two estimators of dissipation based on the infimum statistics of  currents.    
\begin{enumerate}
\item The {\it infimum ratio} $\hat{s}_{\rm inf}$:   defined in Eq.~(\ref{eq:sinf}),  the infimum ratio applies to generic currents $J$ of the form  Eq.~(\ref{eq:JGeneral}).     The infimum ratio is related to the first-passage ratio  $\hat{s}_{\rm FPR}$ through 
\begin{equation}
        \hat{s}_{\rm inf}(\ell_-) = \lim_{\ell_+\rightarrow \infty}\hat{s}_{\rm FPR}(\ell_-,\ell_+),
\end{equation}
and therefore it  inherits the properties of $\hat{s}_{\rm FPR}$, viz., 
\begin{equation}
\lim_{\ell\rightarrow \infty}\hat{s}_{\rm inf}(\ell)  \leq \dot{s},  \label{eq:ineqS}
\end{equation}
and 
\begin{equation}
\lim_{\ell \rightarrow \infty}\hat{s}_{\rm inf}(\ell) = \dot{s}  \label{eq:eqS}
\end{equation}
when $J$ is proportional to $S$.   Moreover, for currents $J$ that are  edge currents $J_{x\rightarrow y}$, it follows from Eq.~(\ref{eq:FPRBiasx}) that 
\begin{equation}
 \lim_{\ell \rightarrow \infty}\hat{s}_{\rm inf}(\ell) = a^\ast_{x\rightarrow y} \overline{j}_{x\rightarrow y}.  \label{eq:SInfxB}
\end{equation}
The drawback of $\hat{s}_{\rm inf}$ is that the Eqs.~(\ref{eq:ineqS})-(\ref{eq:SInfxB})  hold asymptotically   in the limit of  large thresholds $\ell$.    However, for edge currents we can resolve this infinite threshold problem with the next estimator that we discuss.     

\item The {\it modified infimum ratio} $\doublehat{s}_{\rm inf}$:  this ratio, defined in Eq.~(\ref{eq:modified}), applies to edge currents $J_{x\rightarrow y }$.   It follows readily from the definitions  Eqs.~(\ref{eq:sinf})and (\ref{eq:modified}) and  the result Eq.~(\ref{eq:distriExp}) that 
\begin{equation}
    \doublehat{s}_{\rm inf} = a^\ast_{x\rightarrow y} \overline{j}_{x\rightarrow y},
\end{equation}
and hence the modified infimum ratio at finite values of $\ell$ equals the infimum ratio $\hat{s}_{\rm inf}(\ell)$ in the limit of large $\ell$.    Although the modified infimum ratio does not apply to generic currents $J$, it has the advantage  that it uses $p_-(
\ell)$ at finite values of $\ell$, and hence it resolves the infinite threshold problem of the estimators $\hat{s}_{\rm inf}$ and $\hat{s}_{\rm FPR}$.    

\end{enumerate}

\begin{figure*}[t!]\centering
 \includegraphics[width=0.4\textwidth]{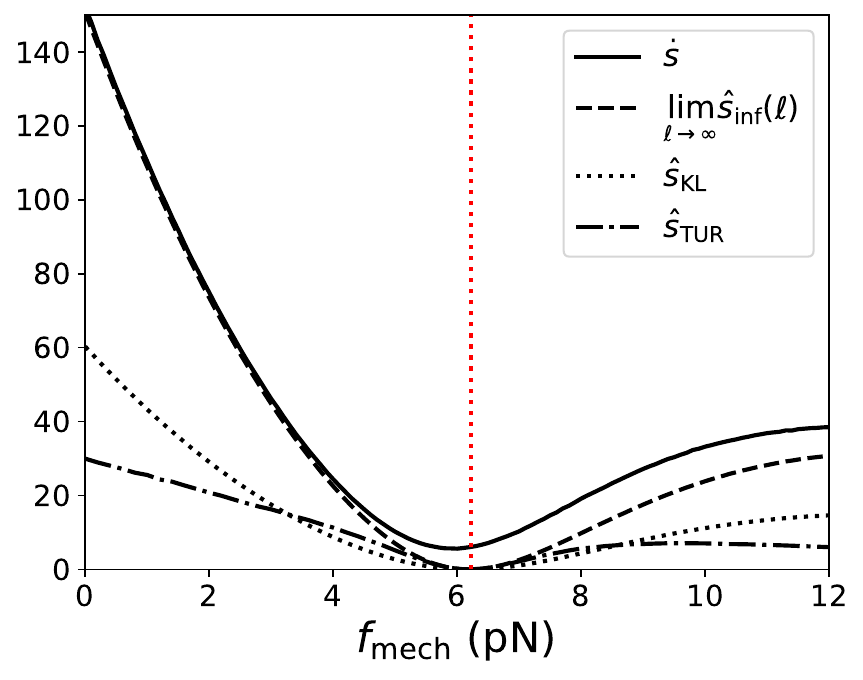}
\includegraphics[width=0.4\textwidth]{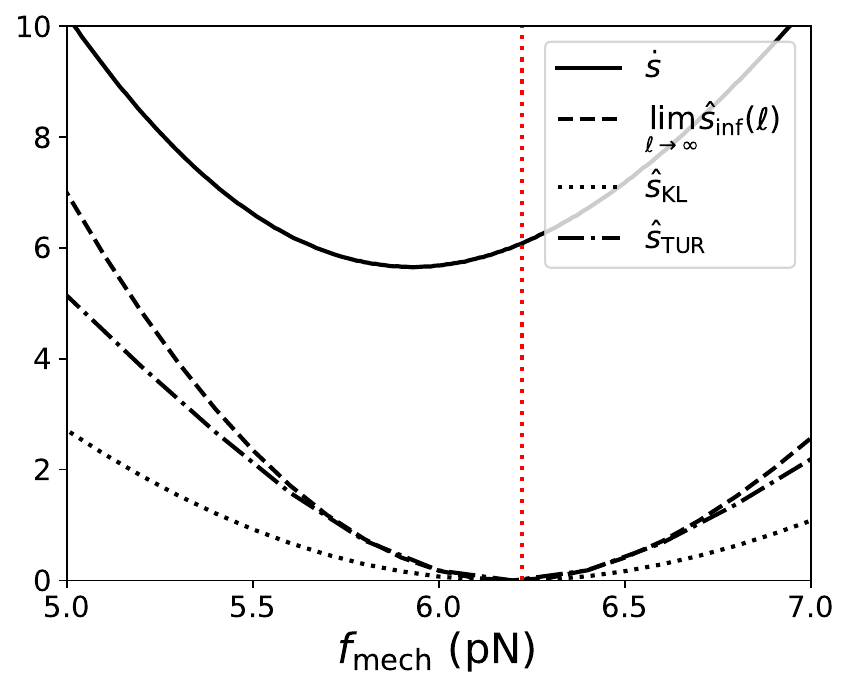}
 \caption{Three estimators of dissipation, viz., $\hat{s}_{\rm KL}$, $\hat{s}_{\rm TUR}$, and $\doublehat{s}_{\rm inf} = \lim_{\ell\rightarrow\infty}\hat{s}_{\rm inf}(\ell)$, are evaluated for the position  $J_{\rm pos}$ of a two-headed molecular motor.   The estimators are plotted as a function of the mechanical force $f_{\rm  mech}$ together with the   entropy production rate $\dot{s}$.  The dynamics of the molecular motor is determined by the model in Sec.~\ref{sec:VIA}  and the parameters used are identical as in Fig.~\ref{fig:results}.   The right figure is a closeup of the left figure around the stalled, nonequilibrium state, denoted by the vertical dotted line.   All the rates are reported in ${\rm s}^{-1}$ and $\hat{s}_{\rm TUR}$ is evaluated at $t = 100s$ (other values of $t$ give similar results).   The $\doublehat{s}_{\rm inf} = \lim_{\ell\rightarrow\infty}\hat{s}_{\rm inf}(\ell)$ is a plot of   Eqs.~(\ref{eq:sinf1})-(\ref{eq:j25exp}).} \label{fig:5}
\end{figure*}

\subsection{Estimation of dissipation in a molecular motor model}\label{sec:83}
We show the different estimators at work on the paradigmatic example of a molecular motor that is bound to a biofilament.   We consider  an experimenter that measures the position $J_{\rm pos}$ of the  two-headed molecular motor, as defined in Sec.~\ref{sec:VIA}.     To this aim, the experimenter uses the three estimators $\hat{s}_{\rm KL}$, $\hat{s}_{\rm TUR}$, and $\doublehat{s}_{\rm inf} = \lim_{\ell\rightarrow \infty}\hat{s}_{\rm inf}(\ell)$.    

Since $J_{\rm pos} = J_{2\rightarrow 5}$, it holds that 
\begin{equation}
\doublehat{s}_{\rm inf}(\ell) = a^\ast_{2\rightarrow 5}\overline{j}_{2\rightarrow 5}, \label{eq:sinf1}
\end{equation}
for $\ell\in \mathbb{N}$.   Using  in Eq.~(\ref{eq:aast25x})  the explicit expressions for $p^{2,5}_{\rm ss}(x)$ reported in Appendix~\ref{app:F}, we obtain for the effective affinity the formula
\begin{eqnarray}
a^\ast_{2\rightarrow 5} &=& \ln \frac{k_{1\rightarrow 2} k_{2\rightarrow 3} k_{3\rightarrow 4} +k_{3\rightarrow 2} k_{4\rightarrow 3} k_{5\rightarrow 4}}{k_{1\rightarrow 2} k_{2\rightarrow 3} k_{3\rightarrow 4} +k_{2\rightarrow1} k_{3\rightarrow 2} k_{4\rightarrow 3}}
\nonumber\\ 
&& + \frac{1}{2}\ln \frac{k_{2\rightarrow 1}(0)}{k_{5\rightarrow 4}(0)}- \frac{f_{\rm mech}\delta}{\mathsf{T}_{\rm env}}  \label{eq:aAst25}  ,
\end{eqnarray}
where we omitted the explicit dependence of the rates on $f_{\rm mech}$ in the first term.   Notice that the average current  $\overline{j}_{2\rightarrow 5}$ is given by 
\begin{equation}
    \overline{j}_{2\rightarrow 5} = p_{\rm ss}(2)k_{2\rightarrow 5} - p_{\rm ss}(5)k_{5\rightarrow 2}.
\label{eq:j25exp}\end{equation}

Figure~\ref{fig:5} shows the quality of  the three estimators $\doublehat{s}_{\rm inf}$, $\hat{s}_{\rm KL}$, and $\hat{s}_{\rm TUR}$ when they are evaluated on the position $J=J_{\rm pos}$ of the molecular motor.   Remarkably, the estimator $\doublehat{s}_{\rm inf}=\lim_{\ell\rightarrow\infty}\hat{s}_{\rm inf}(\ell)$ based on the extreme value statistics  of $J_{\rm pos}$ captures a significant fraction of the dissipation, even in regimes far from thermal equilibrium where both the the thermodynamic uncertainty relation $\hat{s}_{\rm TUR}$ and the Kullback-Leibler divergence $\hat{s}_{\rm KL}$ capture a small proportion of the dissipation.        Indeed, as discussed in Ref.~\cite{neri2}, the thermodynamic uncertainty relation captures a negligible fraction of $\dot{s}$ in regimes far from thermal equilibrium, and $\hat{s}_{\rm KL}$ is strongly biased when the statistics of the current are non-Markovian.    However, as shown in Fig.~\ref{fig:5}, in contrast with $\hat{s}_{\rm TUR}$ and $\hat{s}_{\rm KL}$,  the estimator $\doublehat{s}_{\rm inf}$  accurately  estimates  entropy production rates  far from thermal equilibrium.  A notable exception is when the process is near a nonequilibrium stalled state (the vertical dotted line in Fig.~\ref{fig:5}), in which case none of the estimators capture the dissipation in the process.

As discussed, none of the estimators considered  in this paper  capture dissipation of the process near the stalling state.     However, it should be emphasized that the estimator $\hat{s}_{\rm KL}$ is a  crude approximation of the Kullback-Leibler divergence Eq.~(\ref{eq:KLJ}), as it ignores nonMarkovian correlations in the trajectories of $J$.    As shown in Refs.~\cite{martinez2019inferring, harunari2022learn, van2022thermodynamic}, by considering nonMarkovian effects, such as the  statistics of the  transition times along the edge,  a fraction of the dissipation can be estimated, even at the stalling state.

\section{Discussion}\label{sec:VII}
We have  shown that   the probability mass functions of infima of  empirical, integrated, edge currents in nonequilibrium, stationary states of Markov jump processes are those of a geometric distribution.    The  geometric distribution is determined by two parameters, viz.,  the effective affinity $a^\ast_{x\rightarrow y}$, given by Eq.~(\ref{eq:effA}), and the probability $p_{\rm esc}$ that the infimum equals zero, determined by Eqs.~(\ref{eq:cumula2})-(\ref{eq:pescFinal}).     In general, the latter probability does not admit a simple expression in terms of  $a^\ast_{x\rightarrow y}$, except when the process starts in the  source state $x$ of the observed transition.

The result Eq.~(\ref{eq:distriExp}) implies that the  probability mass function of $J_{x\rightarrow y}^{\rm inf}$ is that of   a  geometric distribution, independent of the underlying model.  As we elaborate in Appendix~\ref{app:H},   this property is specific for edge currents, and hence can be used  to test whether an observed current $J$ in a process $X$ --- we assume here that the observer can measure $J$ but not $X$   --- is an edge current.    Similar tests of transition specificity have been proposed in Ref.~\cite{van2022thermodynamic, harunari2022learn}.

To derive the main results, we have identified the set of martingales $M_{x\rightarrow y}$ (see Eq.~(\ref{eq:Mxy})) associated with the  edge currents  $J_{x\rightarrow y}$.    The martingales $M_{x\rightarrow y}$ are Radon-Nikodym derivative processes, similar to other martingales studied in physics, such as  the exponentiated negative entropy production \cite{chetrite2011two, neri2017statistics, neri2019integral, PhysRevLett.122.220602, PhysRevLett.124.040601, ge2021martingale,PhysRevLett.126.080603,  manzano2022quantum} and the exponentiated housekeeping heat \cite{chetrite2019martingale}.   However, the conjugate probability measure defining the martingales $M_{x\rightarrow y}$ is not simply related to time-reversal (see Eq.~(\ref{eq:radon2x})), as is the case for the entropy production.    It will be interesting to find
other examples of martingales in nonequilibrium physics, in particular, in physical contexts that we have not considered before.     In this regard note  the recent works \cite{PhysRevE.101.062139, PhysRevE.105.044146}, which show  that  the mean equilibrium  value of an unquenched spin in a fully connected spin model under progressive quenching is a martingale. 

   A marginal observer that only observes a current $J$  can estimate the  average rate of entropy production  $\dot{s}$ from the extreme value statistics of a current $J$ through the estimator $\hat{s}_{\rm inf}(\ell)$ in the limit of large $\ell$ (see Eq.~(\ref{eq:sinf}) for a definition of $\hat{s}_{\rm inf}$); this estimator is smaller or equal than $\dot{s}$ and is  equal to $\dot{s}$ when the observed current is proportional to the entropy production $S$~\cite{roldan2015decision, neri2021universal, neri2}.   In this paper, we have shown that for   the particular case when the observed current equals an edge current, i.e., $J=J_{x\rightarrow y}$,  it holds that $\hat{s}_{\rm inf} = a^\ast_{x\rightarrow y}\overline{j}_{x\rightarrow y}$, consistent with the thermodynamic interpretation of $a^\ast_{x\rightarrow y}$ as an effective affinity, see  Refs.~\cite{polettini2017effective, bisker2017hierarchical, polettini2019effective}. 

   The estimator $\hat{s}_{\rm inf}(\ell)$ lower bounds the rate of dissipation in the limit of large $\ell$.   However, in this limit 
   \begin{equation}
   p_-(\ell) = \exp\left(-a^\ast \ell [(1+o(\ell)]\right), 
   \end{equation}
   where the prefactor $a^\ast>0$ is the effective affinity, and therefore the number of samples $n_s\sim 1/p_-$ required to estimate $p_-$ increases exponentially in $\ell$, see~Ref.~\cite{neri2}, which we have called the infinite threshold problem.   In this Paper,  we have shown that for edge currents the average rate of dissipation can be estimated from the extreme value statistics of a current at finite thresholds $\ell$ through the estimator $\doublehat{s}_{\rm inf}$ (see Eq.~(\ref{eq:modified})).      This resolves, for the case of edge currents, the problem with infinite thresholds  when estimating dissipation based on extreme value statistics, which is a special case of the  first passage problem considered in Refs.~\cite{roldan2015decision, neri2021universal, neri2}.     This raises the interesting question  whether the infinite threshold problem for  estimators based on first passage processes  can also be resolved for   currents that are not edge currents.

\section*{Acknowledgements}

IN thanks C.~Hyeon for a useful email communication and A.~Raghu for carefully reading the manuscript.     The research was supported by the National Research Fund Luxembourg (project  CORE ThermoComp C17/MS/11696700) and by the European Research Council, project NanoThermo (ERC-2015-CoG Agreement No. 681456).

\appendix 

\section{Radon-Nikodym derivative processes are martingales}\label{App:A}
We show that Radon-Nikodym derivative processes (as defined in Eq.~(\ref{eq:Radonx}) or(\ref{eq:Radon})) are martingales.  Since the conditions (i) and (ii) of the martingale definition in Sec.~\ref{sec:martingales} are immediate, we focus on demonstrating the condition (iii), given by Eq.~(\ref{eq:MDrift}).

\subsection{Radon-Nikodym derivative processes as conditional expectations}  
Consider the filtered probability space $(\Omega,\mathscr{F},\left\{\mathscr{F}_s\right\}_{s\in\mathbb{R}^+},\mathbb{P})$  generated by the process $X$.  Let $\mathbb{Q}$ be a second probability measure that is locally, absolutely continuous with respect to $\mathbb{P}$; note that we have dropped the $\pinit$ and $\qinit$ from $\mathbb{Q}$ and $\mathbb{P}$ as the arguments presented are general and not restricted to Markov jump processes.    

We  consider  the Radon-Nikodym derivative process~\cite{liptser1977statistics}
\begin{equation}
R(s) :=     \frac{{\rm d}\mathbb{Q}[X^s_0]}{{\rm d}\mathbb{P}[X^s_0]}, \label{eq:defR}
\end{equation} 
with $s\in[0,t]$, 
and aim to show that  
\begin{equation}
R(s) = \langle R(t)| X^s_0\rangle , \quad \forall s\in [0,t],       \label{eq:R}
\end{equation}
holds $\mathbb{P}$-almost surely, 
where $\langle \cdot|X^s_0\rangle$ is the conditional expectation with respect to the sub-$\sigma$-algebra $\mathscr{F}_s$ generated by the trajectory $X^s_0$.    

To show that Eq.~(\ref{eq:R}) holds, we first use the definition of the conditional expectation $\langle R(t)| X^s_0\rangle$, viz., $\langle R(t)| X^s_0\rangle$ is   a random variable defined  on $(\Omega,\mathscr{F}_s)$~\cite{liptser1977statistics} for which 
\begin{equation}
    \int_{\Phi} \langle R(t)|X^s_0\rangle {\rm d} \mathbb{P}[X^s_0]   =  \int_{\Phi} R(t)  {\rm d} \mathbb{P}[X^t_0] \label{eq:integral}
\end{equation}
holds for all $\Phi\in \mathscr{F}_s$.    Subsequently, we use that $R(t)$ is the Radon-Nikodym derivative (\ref{eq:defR}) to write   the right-hand side of Eq.~(\ref{eq:integral}) as
\begin{equation}
\int_{\Phi} R(t)  {\rm d} \mathbb{P}[X^t_0] = \int_{\Phi}    {\rm d} \mathbb{Q}[X^t_0]. \label{eq:1x}
\end{equation}
Marginalising the latter distribution leads to 
\begin{equation}
\int_{\Phi}    {\rm d} \mathbb{Q}[X^t_0] = \int_{\Phi} {\rm d} \mathbb{Q}[X^s_0], \label{eq:2x}
\end{equation}
and using that  $R(s)$ is the Radon-Nikodym derivative (\ref{eq:defR}), we obtain  \begin{equation}
\int_{\Phi}    {\rm d} \mathbb{Q}[X^s_0] = \int_{\Phi}   R(s)  {\rm d} \mathbb{P}[X^s_0].  \label{eq:3x}
\end{equation}
Equations (\ref{eq:integral})-(\ref{eq:3x}) imply that Eq.~(\ref{eq:R}) holds $\mathbb{P}$-almost surely, which we were meant to show.

\subsection{Martingale property of $R$}
The martingale property (iii), given by Eq.~(\ref{eq:MDrift}),  for  $R(t)$ is   a direct consequence of  Eq.~(\ref{eq:R}) and the tower property 
\begin{equation}
    R(s')  =   \langle R(t)|  X^{s'}_0\rangle =  \langle  \langle R(t)| X^s_0\rangle | X^{s'}_0\rangle = \langle  R(s)|X^{s'}_0\rangle ,\quad {\rm with} \quad  0\leq s'\leq s\leq t, 
\end{equation}
that holds for conditional expectations of a random variable \cite{williams1991probability}.        

\section{Derivation of Eq.~(\ref{eq:radon2x})}\label{app:BBis}
We show that the martingale $M_{x\rightarrow y}$, given by Eq.~(\ref{eq:Mxy}), is the Radon-Nikodym derivative process Eq.~(\ref{eq:radon2x}), where $(X,\mathbb{R}_{{\rm ss}})$ is the Markov jump process with rates $m_{x\rightarrow y}$ given by Eq.~(\ref{homx}).  To this aim, we use the formula (\ref{eq:Radon3}), which is valid when the two conditions Eqs.~(\ref{eq:exitId}) and (\ref{eq:rateId}) hold. 
 
First, we verify (\ref{eq:exitId}), i.e., we verify that the exit rates 
\begin{equation}
    \sum_{v\in \mathcal{X};(v\neq u)}k_{u\rightarrow v} =    \sum_{v\in \mathcal{X};(v\neq u)}m_{u\rightarrow v}. \label{eq:tocheck}
\end{equation}
Using the definition  Eq.~(\ref{homx}) for the rates $m_{u\rightarrow v}$ together with the fact that, by definition,  $q_{\rm ss}$ satisfies the stationary conditions 
\begin{equation}
     \sum_{v\in \mathcal{X};(v\neq u)}q_{\rm ss}(v)\ell_{v\rightarrow u} = q_{\rm ss}(u)  \sum_{v\in \mathcal{X};(v\neq u)}\ell_{u\rightarrow v}
\end{equation}
for the Markov jump process $(X,\mathbb{Q}_{\rm ss})$  with rates $\ell_{u\rightarrow v}$  given by Eq.~(\ref{hom}), we recover (\ref{eq:tocheck}).   

Second, we verify (\ref{eq:rateId}), which follows readily from the definition of the rates (\ref{homx}). 

Hence we can use Eq.~(\ref{eq:Radon3}), for the Radon-Nikodym derivative of $\mathbb{R}_{{\rm ss}}$ with respect to $\mathbb{P}_{\rm ss}$, to  obtain 
\begin{equation}
 \frac{{\rm d}\mathbb{R}_{{\rm ss}}[X^t_0]}{{\rm d}\mathbb{P}_{{\rm ss}}[X^t_0]}   = \frac{q_{\rm ss}(X(0))}{p_{\rm ss}(X(0))} \exp\left(\frac{1}{2} \sum_{(u, v)\in \mathcal{E}} J_{u\rightarrow v}(\omega,t) \ln \frac{m_{u\rightarrow v}}{k_{u\rightarrow v}} \right). \label{eq:120}
\end{equation}
Consequently, using the definition (\ref{homx}) in (\ref{eq:120}), we obtain 
\begin{equation}
 \frac{{\rm d}\mathbb{R}_{{\rm ss}}[X^t_0]}{{\rm d}\mathbb{P}_{\rm ss}[X^t_0]}   = \frac{q_{\rm ss}(X(0))}{p_{\rm ss}(X(0))} \exp\left( \frac{1}{2}\sum_{(u, v)\in \mathcal{E}} J_{u\rightarrow v}(\omega,t) \ln \frac{q_{\rm ss}(v)p^{x,y}_{\rm ss}(u)}{q_{\rm ss}(u)p^{x,y}_{\rm ss}(v)}   +  J_{x\rightarrow y} \ln \frac{p^{x,y}_{\rm ss}(y)k_{y\rightarrow x}}{p^{x,y}_{\rm ss}(x)k_{x\rightarrow y}}\right), \label{eq:121}
\end{equation}
where the $J_{x\rightarrow y}\ln p^{x,y}_{\rm ss}(x)/p^{x,y}_{\rm ss}(y)$ in the first term of the exponent cancels out with the $J_{x\rightarrow y} \ln p^{x,y}_{\rm ss}(y)/p^{x,y}_{\rm ss}(x)$ in the second term of the exponent.  In addition, identifying 
\begin{equation}
   q_{\rm ss}(X(0)) \exp\left(\frac{1}{2} \sum_{(u, v)\in \mathcal{E}} J_{u\rightarrow v}(\omega,t) \ln \frac{q_{\rm ss}(v)}{q_{\rm ss}(u)} \right) = q_{\rm ss}(X(t))
\end{equation}
and 
\begin{equation}
  p^{x,y}_{\rm ss}(X(t))  \exp\left(\frac{1}{2}  \sum_{(u, v)\in \mathcal{E}} J_{u\rightarrow v}(\omega,t) \ln \frac{p^{x,y}_{\rm ss}(u)}{p^{x,y}_{\rm ss}(v)} \right) = p^{x,y}_{\rm ss}(X(0)),
\end{equation}
in (\ref{eq:121}), we obtain 
\begin{equation}
 \frac{{\rm d}\mathbb{R}_{{\rm ss}}[X^t_0]}{{\rm d}\mathbb{P}_{\rm ss}[X^t_0]}   =\frac{p_{\rm ss}^{x,y}(X(0))q_{\rm ss}(X(t))}{p_{\rm ss}^{x,y}(X(t))p_{\rm ss}(X(0))} e^{- a^\ast_{x\rightarrow y}J_{x\rightarrow y}(t)} = M_{x\rightarrow y}(t) ,
\end{equation} 
and thus according to (\ref{eq:Mxy}) we find  (\ref{eq:radon2x}), which is our desired result.

\section{Microscopic and effective affinities in unicyclic systems}\label{App:Cb}

We compare the microscopic affinities $a_{x\rightarrow y}$, as defined by Eq.~(\ref{eq:a}), with the effective affinities $a^\ast_{x\rightarrow y}$, as defined by Eq.~(\ref{eq:effA}), in  unicyclic systems.    In particular, we consider systems described by the following Master equation, 
\begin{equation}
\partial_t p(u;t) = p(u+1;t)k_{u+1\rightarrow u} +  p(u-1;t)k_{u-1\rightarrow u} - (k_{u\rightarrow u+1} + k_{u\rightarrow u-1})p(u;t)  \label{eq:uni}
\end{equation}
where $u\in \mathcal{X} = \left\{1,2,\ldots,\ell\right\}$, and in Eq.~(\ref{eq:uni}) it should be understood that $0=\ell$ and $\ell+1 = 1$.  
  
At stationarity, $\partial_t p(u;t) = 0$, such that the  edge currents given by Eq.~(\ref{eq:edgeMarkovCurrent}) obey
\begin{equation}
\overline{j}_{u-1\rightarrow u} =\overline{j}_{u\rightarrow u+1} = \overline{j} \label{eq:JC}
\end{equation}
for all $u\in\mathcal{X}$.  Using Eq.~(\ref{eq:JC}) in Eq.~(\ref{eq:sdotDef}), we obtain for the rate of  dissipation,
\begin{equation}
\dot{s} = a \overline{j},
\end{equation}
where $a$ is identified as the "true", macroscopic affinity
\begin{equation}
a =  \sum^{\ell}_{u=1}a_{u\rightarrow u+1}.
\end{equation}
Note that the macroscopic affinity $a$ is the sum of all microscopic affinities $a_{u\rightarrow v}$.  Hence,  in general, the  microscopic affinities  contribute a small part of the total affinity.   For example, when $k_{u\rightarrow u+1} = k_+$ for all $u\in\mathcal{X}$, and $k_{u\rightarrow u-1} = k_-$ for all $u\in\mathcal{X}$,  then 
\begin{equation}
a_{u\rightarrow u+1} = \ln \frac{k_+}{k_-} 
\end{equation}
 and 
\begin{equation}
a = \ell \ln \frac{k_+}{k_-} . 
\end{equation} 

Let us now determine the effective affinities $a^\ast_{x\rightarrow y}$ of unicyclic Markov processes described by Eq.~(\ref{eq:uni}).  To determine $a^\ast_{x\rightarrow y}$, we need to determine the values $p^{x,x+1}_{\rm ss}(x)$  and $p^{x,x+1}_{\rm ss}(x+1)$ of the stationary distributions $p^{x,x+1}_{\rm ss}(u)$ solving
\begin{equation}
0 = p^{x,x+1}_{\rm ss}(u+1)k_{u+1\rightarrow u} +  p^{x,x+1}_{\rm ss}(u-1)k_{u-1\rightarrow u} - (k_{u\rightarrow u+1} + k_{u\rightarrow u-1})p^{x,x+1}_{\rm ss}(u)  \label{eq:uni2}
\end{equation}
for all $u\in \mathcal{X}\setminus \left\{x,x+1\right\}$, 
\begin{equation}
 k_{x\rightarrow x-1}\: p^{x,x+1}_{\rm ss}(x) =  p^{x,x+1}_{\rm ss}(x-1)k_{x-1\rightarrow x}, \label{eq:uni3}
\end{equation}
and 
\begin{equation}
k_{x+1\rightarrow x+2}\: p^{x,x+1}_{\rm ss}(x+1)  = p^{x,x+1}_{\rm ss}(x+2)k_{x+2\rightarrow x+1}   . \label{eq:uni4}
\end{equation}
Solving the Eqs.~(\ref{eq:uni2}-\ref{eq:uni4}), we obtain 
\begin{equation}
p^{x,x+1}_{\rm ss}(x) = p_0 \prod^{\ell}_{u=1;u\neq x}\frac{k_{u\rightarrow u+1}}{k_{u+1\rightarrow u}} \label{eq:pxx1}
\end{equation}
and 
\begin{equation}
p^{x,x+1}_{\rm ss}(x+1) = p_0, \label{eq:pxx2}
\end{equation}
where $p_0$ is a normalisation constant.  Substitution of Eqs.~(\ref{eq:pxx1}) and (\ref{eq:pxx2}) in the definition (\ref{eq:effA}) of  $a^\ast_{x\rightarrow x+1}$ yields 
\begin{equation}
a^\ast_{x\rightarrow x+1} =  \sum^{\ell}_{u=1}a_{u\rightarrow u+1} = a. 
\end{equation}
Hence, the effective affinity in a unicyclic system equals the macroscopic affinity.

\section{Derivation of $a^\ast_{x\rightarrow y}\langle J_{x\rightarrow y}(t)\rangle _{\rm ss} \geq 0$}\label{App:B}
We use the $\mathbb{P}_{\rm ss}$-martingale $M_{x\rightarrow y}(t)$, given by  Eq.~(\ref{eq:Mxy}) to show that $a^\ast_{x\rightarrow y}\langle J_{x\rightarrow y}(t)\rangle_{\rm ss} \geq 0$.  

 Indeed, since  $M_{x\rightarrow y}(t)$ is a $\mathbb{P}_{\rm ss}$-martingale, it holds that 
\begin{equation}
\langle  M_{x\rightarrow y}(t)\rangle_{\rm ss} = \langle M_{x\rightarrow y}(0) \rangle_{\rm ss}  = 1 . 
\end{equation}
In addition, since 
\begin{equation}
M_{x\rightarrow y}(t) = e^{-a^\ast J_{x\rightarrow y}(t) + O_t(1)},
\end{equation}
where the big-O notation $O_t(1)$ denotes an arbitrary function of $t$ that is bounded, it holds that 
\begin{equation}
\langle  M_{x\rightarrow y}(t)\rangle_{\rm ss} = \langle e^{-a^\ast_{x\rightarrow y} J_{x\rightarrow y}(t) + O_t(1)} \rangle_{\rm ss} = 1.  
\end{equation}
Applying Jensen's inequality  
\begin{equation}
\langle e^{-a^\ast_{x\rightarrow y} J_{x\rightarrow y}(t) + O_t(1)}\rangle_{\rm ss} \geq e^{-a^\ast_{x\rightarrow y} \langle J_{x\rightarrow y}(t)\rangle_{\rm ss}  + O_t(1)} \label{eq:Jensen} \end{equation}
and using  
\begin{equation}
    \langle J_{x\rightarrow y}(t)\rangle_{\rm ss} = \overline{j}_{x\rightarrow y}t, 
\end{equation}
with $\overline{j}_{x\rightarrow y}\in \mathbb{R}$ the average current,
we obtain 
\begin{equation}
a^\ast_{x\rightarrow y} \overline{j}_{x\rightarrow y}t  + O_t(1) \geq 0. 
\end{equation} 
Since $O_t(1)/t\rightarrow 0$, it holds that 
\begin{equation}
a^\ast_{x\rightarrow y}\langle J_{x\rightarrow y}(t)\rangle_{\rm ss} \geq 0. \label{eq:astxyP}
\end{equation} 
 
According to Eq.~(\ref{eq:astxyP}), $\langle J_{x\rightarrow y}(t)\rangle_{\rm ss}$  changes sign when $a^\ast_{x\rightarrow y}=0$.   Hence, $a^\ast_{x\rightarrow y}=0$    if and only if $\langle J_{x\rightarrow y}(t)\rangle_{\rm ss} = 0$.

  \section{Equilibrium distribution in the six state model}\label{App:C}
  
The equilibrium state of the six-state model defined in Sec.~\ref{sec:kinesin} is 
\begin{eqnarray}
p_{\rm eq}(1)=\frac{1}{\mathcal{N}}, \quad  p_{\rm eq}(2) =\frac{1}{\mathcal{N}}\sqrt{\frac{k_{5\rightarrow 4}(0)}{k_{2\rightarrow 1}(0)}}\frac{k_{4\rightarrow 3}(0) }{k_{3\rightarrow 4}(0)}  \frac{k_{3\rightarrow 2}(0) }{k_{2\rightarrow 3}(0)},   \quad p_{\rm eq}(3) =\frac{1}{\mathcal{N}} \sqrt{\frac{k_{5\rightarrow 4}(0)}{k_{2\rightarrow 1}(0)}}\frac{k_{4\rightarrow 3}(0) }{k_{3\rightarrow 4}(0)},  \nonumber \\ 
 p_{\rm eq}(4) = \frac{1}{\mathcal{N}}\sqrt{\frac{k_{5\rightarrow 4}(0)}{k_{2\rightarrow 1}(0)}},  \quad p_{\rm eq}(5) = \frac{1}{\mathcal{N}}\frac{k_{4\rightarrow 3}(0) }{k_{3\rightarrow 4}(0)}  \frac{k_{3\rightarrow 2}(0) }{k_{2\rightarrow 3}(0)},  \quad  p_{\rm eq}(6) = \frac{1}{\mathcal{N}}\frac{k_{4\rightarrow 3}(0) }{k_{3\rightarrow 4}(0) }, \nonumber\\
\end{eqnarray}
where the normalisation constant is 
\begin{eqnarray}
\mathcal{N} = 1 + \sqrt{\frac{k_{5\rightarrow 4}(0)}{k_{2\rightarrow 1}(0)}}\frac{k_{4\rightarrow 3}(0) }{k_{3\rightarrow 4}(0)}  \frac{k_{3\rightarrow 2}(0) }{k_{2\rightarrow 3}(0)}  +  \sqrt{\frac{k_{5\rightarrow 4}(0)}{k_{2\rightarrow 1}(0)}}\frac{k_{4\rightarrow 3}(0) }{k_{3\rightarrow 4}(0)}  
\nonumber \\ 
+\sqrt{\frac{k_{5\rightarrow 4}(0)}{k_{2\rightarrow 1}(0)}} 
+ \frac{k_{4\rightarrow 3}(0) }{k_{3\rightarrow 4}(0)}  \frac{k_{3\rightarrow 2}(0) }{k_{2\rightarrow 3}(0)} + \frac{k_{4\rightarrow 3}(0) }{k_{3\rightarrow 4}(0) }.
\end{eqnarray}
\section{Parameters  for the six-state model of Kinesin-1}\label{app:C}
We specify the parameters that we use in Sec.~\ref{VI:C} for the  six-state model for two-headed molecular motors, as visualised in Fig.~\ref{fig:examples}.     

We use the parametrisation  of the rates given by Eqs.~(\ref{eq:param1}), (\ref{eq:param2}),  (\ref{eq:param3}), (\ref{eq:atp1}), and (\ref{eq:atp2}).     We set the parameters $k_{i\rightarrow j}(0)$, $\theta$, and $\chi_{ij}$ to the same values  as those given   in Refs.~\cite{hwang2018energetic}, which were obtained by fitting    single molecule motility data of Kinesin-1 to the six-state model, except for the parameter $k_{5\rightarrow 2}(0)$, which we set to (\ref{eq:K52}) such that the model obeys detailed balance when $f_{\rm mech}$ and  $[\rm ATP]$ are given by Eq.~(\ref{eq:eqcond}).   Note that from the pragmatic point of view of modelling the dynamics of Kinesin-1, this does not make much of a difference, as in our case $k_{5\rightarrow 2}(0) \approx 1.13 \: {\rm s}^{-1}$ while in Ref.~\cite{hwang2018energetic} $k_{5\rightarrow 2}(0) = 1.1 \: {\rm s}^{-1}$.   However, from a theoretical point of view, it is desirable to have an equilibrium point in the model as it allows us to study the properties of the system near equilibrium.     

Concretely, we set $\theta = 0.61$, $\chi_{12} = 0.15$, $\chi_{56} = 0.0015$, $\chi_{61} = 0.11$, $k_{2\rightarrow 1}(0) = 4200 \: s^{-1}$, $k_{2\rightarrow5}(0) = 1.6 \times 10^6 \:  s^{-1}$, $k_{5\rightarrow2}(0) = 1.1 \:  s^{-1}$,$k_{5\rightarrow6}(0)=190 \:  s^{-1}$, $k_{6\rightarrow5}(0)=10 \:  s^{-1}$, $k_{6\rightarrow1}(0) = 250 \:  s^{-1}$, $k_{1\rightarrow6}(0) = 230\:   s^{-1}$, $k_{5\rightarrow4}(0)= 2.1 \times 10^{-9} \:  s^{-1}$, and we set  $k_{1\rightarrow 2}(0) = k^{\rm bi}_{1\rightarrow 2}(0)[{\rm ATP}]$ with $k^{\rm bi}_{1\rightarrow 2}(0) = 2.8\:   \mu M^{-1}s^{-1}$.

Note that we have set  $k_{3\rightarrow 2}(0) = k_{6\rightarrow 5}(0)$, $k_{2\rightarrow 3}(0) = k_{5\rightarrow 6}(0)$ $k_{3\rightarrow 4}(0) = k_{6\rightarrow 1}(0)$, $k_{4\rightarrow 3}(0) = k_{1\rightarrow 6}(0)$, $k_{4\rightarrow 5}(0) = k_{1\rightarrow 2}(0) $, $\chi_{23}=\chi_{56}$, $\chi_{34}=\chi_{61}$, and $\chi_{45}=\chi_{12}$. 

The molecular motor step size is set to the length of a tubulin dimer (the subunit that forms the microtubule filament that is the substrate to which kinesin binds),  viz.,  $\delta=8\: {\rm nm}$,  and the environment is set to   room temperature, i.e.,  $\mathsf{T}_{\rm env} = 298\times 1.38\times 10^{-23}\: {\rm J}$.

\section{The stationary distribution of the six-state model in  the absence of the $2\rightarrow 5$ link}\label{app:F}
The stationary distribution $p^{2,5}_{\rm ss}(x)$ of the six-state model, defined in Sec.~\ref{sec:VIA}, in the absence of the $2\rightarrow 5$ and $5\rightarrow2$ transitions is given by:
\begin{eqnarray}
p^{2,5}_{\rm ss}(1) &=& \frac{1}{\mathcal{N}^{2,5}}k_{1\rightarrow 2} k_{2\rightarrow 3} k_{3\rightarrow 4} \left( k_{2\rightarrow 3} k_{3\rightarrow 4} + k_{2\rightarrow 1} \left[k_{3\rightarrow 2} + k_{3\rightarrow 4}\right]\right)  
\nonumber\\ 
&& +\frac{1}{\mathcal{N}^{2,5}}k_{2\rightarrow 1} k_{3\rightarrow 2} k_{4\rightarrow 3} \left(k_{2\rightarrow3} k_{3\rightarrow 4} + \left[k_{3\rightarrow 2} +k_{3\rightarrow 4}\right] k_{5\rightarrow 4}\right) , \nonumber\\
p^{2,5}_{\rm ss}(2) &=& \frac{1}{\mathcal{N}^{2,5}} \left(\left[k_{1\rightarrow 2} \left\{k_{3\rightarrow 2} + k_{3\rightarrow 4}\right\} + k_{3\rightarrow 2} k_{4\rightarrow 3}\right] \left[k_{1\rightarrow 2} k_{2\rightarrow 3} k_{3\rightarrow 4} +k_{3\rightarrow 2} k_{4\rightarrow 3} k_{5\rightarrow 4}\right]\right), \nonumber \\
p^{2,5}_{\rm ss}(3) &=& \frac{1}{\mathcal{N}^{2,5}} k_{1\rightarrow2} k_{2\rightarrow3}^2 k_{3\rightarrow4} \left(k_{1\rightarrow2}+ k_{4\rightarrow3}\right) 
\nonumber\\
&& +\frac{1}{\mathcal{N}^{2,5}}
 k_{4\rightarrow3}k_{5\rightarrow4} \left(k_{1\rightarrow2} k_{2\rightarrow3} \left[k_{3\rightarrow2} + k_{3\rightarrow4}\right] + \left[k_{2\rightarrow1} +k_{2\rightarrow3}\right] k_{3\rightarrow2} k_{4\rightarrow3}\right) ,  \nonumber\\ 
p^{2,5}_{\rm ss}(4) &=& \frac{1}{\mathcal{N}^{2,5}} k_{3\rightarrow2}k_{4\rightarrow3} k_{5\rightarrow4} \left(k_{2\rightarrow3} k_{3\rightarrow4} + k_{2\rightarrow1} \left[k_{3\rightarrow2} + k_{3\rightarrow4}\right]\right) \nonumber\\ 
&&+ \frac{1}{\mathcal{N}^{2,5}}
k_{1\rightarrow2} k_{2\rightarrow3} k_{3\rightarrow4} \left(k_{2\rightarrow3} k_{3\rightarrow4} + k_{5\rightarrow4}\left[k_{3\rightarrow2} +k_{3\rightarrow4}\right] \right),  \nonumber\\ 
  p^{2,5}_{\rm ss}(5) &=&\frac{1}{\mathcal{N}^{2,5}}    \left(\left[k_{1\rightarrow 2} \left\{k_{3\rightarrow 2} + k_{3\rightarrow 4}\right\} + k_{3\rightarrow 2} k_{4\rightarrow 3}\right] \left[k_{1\rightarrow 2} k_{2\rightarrow 3} k_{3\rightarrow 4} +k_{2\rightarrow1} k_{3\rightarrow 2} k_{4\rightarrow 3}\right]\right), \nonumber \\ 
   p^{2,5}_{\rm ss}(6) &=& \frac{1}{\mathcal{N}^{2,5}} \left(k^2_{1\rightarrow2} k^2_{2\rightarrow3} k_{3\rightarrow4} + k_{2\rightarrow1} k_{3\rightarrow2} k^2_{4\rightarrow3} \left[k_{2\rightarrow3} + k_{5\rightarrow4}\right] \right) 
   \nonumber\\ 
   && 
   +\frac{1}{\mathcal{N}^{2,5}}   k_{1\rightarrow2} k_{2\rightarrow3} k_{4\rightarrow3} \left(k_{2\rightarrow3} k_{3\rightarrow4} + k_{2\rightarrow1} \left[k_{3\rightarrow2} + k_{3\rightarrow4}\right]\right),\nonumber
\end{eqnarray}
where 
\begin{eqnarray}
\mathcal{N}^{2,5} &=& 2 k_{1\rightarrow 2}^2 k_{2\rightarrow 3} k_{3\rightarrow 4} \left(k_{2\rightarrow 3} + k_{3\rightarrow 2} + k_{3\rightarrow 4}\right) 
\nonumber\\ 
& +&k_{1\rightarrow 2} \left(k_{3\rightarrow 2} k_{4\rightarrow 3} + k_{2\rightarrow 3} [k_{3\rightarrow 4} + k_{4\rightarrow3}]\right) 
\nonumber\\ 
&& \times
\left(2 k_{2\rightarrow 3} k_{3\rightarrow 4} +k_{2\rightarrow 1} [k_{3\rightarrow 2} + k_{3\rightarrow 4}] + [k_{3\rightarrow 2} + k_{3\rightarrow 4}] k_{5\rightarrow 4}\right)  
\nonumber\\ 
&
+&k_{3\rightarrow 2} k_{4\rightarrow 3} \left(k_{2\rightarrow 1} k_{2\rightarrow 3} k_{3\rightarrow 4} +k_{2\rightarrow 1} [k_{2\rightarrow 3} + k_{3\rightarrow 2}] k_{4\rightarrow 3}  \right.
\nonumber\\
&& \left.
+  k_{5\rightarrow 4}\left[k_{2\rightarrow 3} k_{3\rightarrow 4} + \left\{k_{2\rightarrow 3} + k_{3\rightarrow 2}\right\} k_{4\rightarrow3} +2 k_{2\rightarrow1} \left\{k_{3\rightarrow 2} + k_{3\rightarrow 4} + k_{4\rightarrow 3}\right\}\right]\right) \nonumber\\
\end{eqnarray}
is the normalisation constant.

\section{The modified infimum ratio for empirical, integrated currents that are not edge currents} \label{app:H}
For empirical, integrated, edge currents, $J_{x\rightarrow y},$ it holds that  for all $\ell\in \mathbb{N}$ the modified infimum ratio 
\begin{equation}
\doublehat{s}_{\rm inf} = \ln \frac{p_-(\ell)}{p_-(\ell+1)} = a^{\ast}_{x\rightarrow y}, \label{eq:appH}
\end{equation}
is a constant independent of $\ell$; notice that for $\ell=0$ the modified infimum ratio takes a value different from $a^{\ast}_{x\rightarrow y}$.   Recall that in  Eq.~(\ref{eq:appH}) 
\begin{equation}
p_-(\ell) = \mathbb{P}_{\rm ss}\left(J^{\rm inf}_{x\rightarrow y}\leq -\ell\right).
\end{equation}  

\begin{figure*}[t!]\centering
 \includegraphics[width=0.5\textwidth]{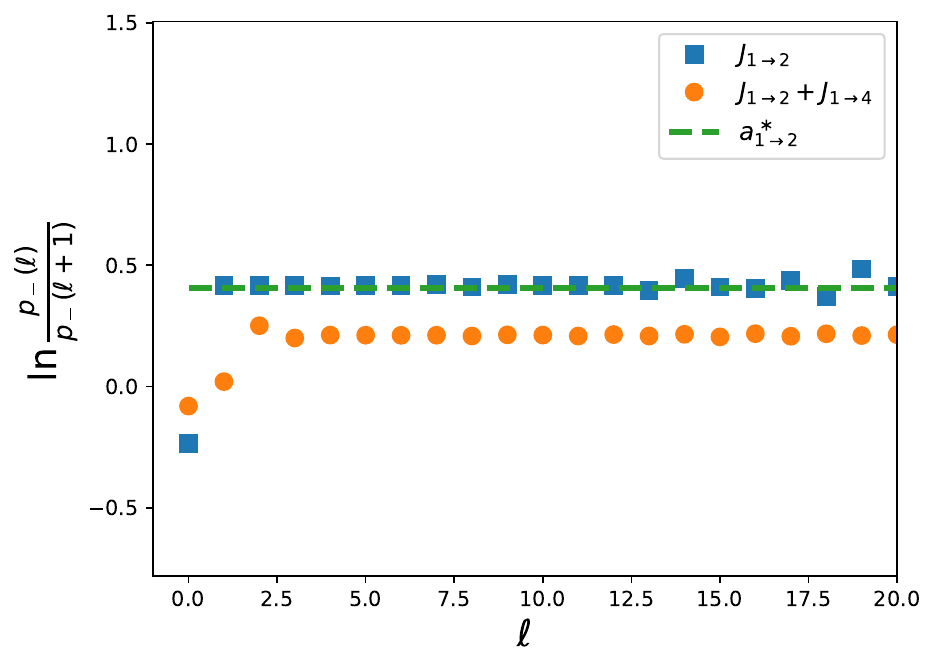}
 \caption{The modified infimum ratio $\doublehat{s}_{\rm inf}$ as a function of $\ell$ for the edge current $J_{1\rightarrow 2}$ (blue squares) and the sum $J_{1\rightarrow 2}+J_{1\rightarrow 4}$ of two edge currents (orange circles) in the stationary Markov jump process with  four states and transition rates given by Eq.~(\ref{eq:kuvStatExamp}) with parameters $c=1.5$ and $b = 1/e$.   Markers are simulation results for $\doublehat{s}_{\rm inf}$  obtained from $1e+7$ simulation runs over a time interval $t\in [0,1e+4]$, which are compared with the effective affinity $a^\ast_{1\rightarrow 2}$ given by Eq.~(\ref{eq:effectAffLNc}).    } \label{fig:6}
\end{figure*}

In the asymptotic limit of large $\ell$, i.e., $\ell\gg 1$, the constancy of $\doublehat{s}_{\rm inf}$ holds  for generic currents $J$.  However, at finite $\ell$, the modified infimum ratio is, in general, not a constant.  Consequently,  the constancy of $\doublehat{s}_{\rm inf}$ for $\ell\in\mathbb{N}$ can be used to  identify whether a current $J$ is an edge current.    

Let us illustrate this on a simple model.   Consider a four state Markov jump process for which $X(t)\in \mathcal{X} = \left\{1,2,3,4\right\}$ and with 
\begin{equation}
k_{u\rightarrow v} = \left\{\begin{array}{ccc} 1, &{\rm if}& u>v , \\  c b^{v-u}, &{\rm if}& u<v ,\end{array}\right. \label{eq:kuvStatExamp}
\end{equation}
where $c,b>0$, and $u,v\in \left\{1,2,3,4\right\}$. 

The stationary distribution of this model is given by
\begin{eqnarray}
p_{\rm ss}(1) &=& \frac{1}{\mathcal{N}}\frac{2+bc+b^2c}{b^3c(1+c+bc+b^2c)},   \quad  p_{\rm ss}(2) = \frac{1}{\mathcal{N}}\frac{2+b+b^2}{b^2(1+c+bc+b^2c)}, \nonumber\\ 
p_{\rm ss}(3) &=& \frac{1}{\mathcal{N}}  \frac{1}{b}, \quad p_{\rm ss}(4) = \frac{1}{\mathcal{N}},
\end{eqnarray}
where $\mathcal{N}$ is a normalisation constant.   

In Fig~\ref{fig:6}, we present numerical simulation results for  the modified infimum ratio $\doublehat{s}_{\rm inf}$ of  the edge current $J_{1\rightarrow 2}$ with effective affinity 
\begin{equation}
a^\ast_{1\rightarrow 2} = \ln  c, \label{eq:effectAffLNc}
\end{equation}
and we also present $\doublehat{s}_{\rm inf}$  evaluated on the infima of the currents $J_{1\rightarrow 2} + J_{1\rightarrow 4}$.  Figure~\ref{fig:6} shows that the modified infimum ratio of the edge current is constant and equal to the effective affinity, while the modified infimum ratios of $J_{1\rightarrow 2} + J_{1\rightarrow 4}$ is  nonconstant for small values of $\ell$ before saturating to a constant value for intermediate values of $\ell$.

\bibliographystyle{SciPost_bibstyle}
\bibliography{bibliography.bib}

\begin{thebibliography}{10}
\providecommand{\url}[1]{\texttt{#1}}
\providecommand{\urlprefix}{URL }
\expandafter\ifx\csname urlstyle\endcsname\relax
  \providecommand{\doi}[1]{doi:\discretionary{}{}{}#1}\else
  \providecommand{\doi}{doi:\discretionary{}{}{}\begingroup
  \urlstyle{rm}\Url}\fi
\providecommand{\eprint}[2][]{\url{#2}}

\bibitem{companion}
M.~Polettini and I.~Neri,
\newblock \emph{Phenomenological boltzmann formula for currents},
\newblock Submitted to SciPost Physics  (2022).

\bibitem{andrieux2007fluctuationa}
D.~Andrieux and P.~Gaspard,
\newblock \emph{Fluctuation theorem for currents and schnakenberg network
  theory},
\newblock Journal of statistical physics \textbf{127}(1), 107 (2007),
\newblock \doi{10.1007/s10955-006-9233-5}.

\bibitem{andrieux2007fluctuation}
D.~Andrieux and P.~Gaspard,
\newblock \emph{A fluctuation theorem for currents and non-linear response
  coefficients},
\newblock Journal of Statistical Mechanics: Theory and Experiment
  \textbf{2007}(02), P02006 (2007),
\newblock \doi{10.1088/1742-5468/2007/02/P02006}.

\bibitem{gaspard2013multivariate}
P.~Gaspard,
\newblock \emph{Multivariate fluctuation relations for currents},
\newblock New Journal of Physics \textbf{15}(11), 115014 (2013),
\newblock \doi{10.1088/1367-2630/15/11/115014}.

\bibitem{pietzonka2015universal}
P.~Pietzonka, A.~C. Barato and U.~Seifert,
\newblock \emph{Universal bounds on current fluctuations},
\newblock Phys. Rev. E \textbf{93}, 052145 (2016),
\newblock \doi{10.1103/PhysRevE.93.052145}.

\bibitem{gingrich2015dissipation}
T.~R. Gingrich, J.~M. Horowitz, N.~Perunov and J.~L. England,
\newblock \emph{Dissipation bounds all steady-state current fluctuations},
\newblock Phys. Rev. Lett. \textbf{116}, 120601 (2016),
\newblock \doi{10.1103/PhysRevLett.116.120601}.

\bibitem{barato2015thermodynamic}
A.~C. Barato and U.~Seifert,
\newblock \emph{Thermodynamic uncertainty relation for biomolecular processes},
\newblock Physical review letters \textbf{114}(15), 158101 (2015),
\newblock \doi{10.1103/PhysRevLett.114.158101}.

\bibitem{PhysRevE.96.012101}
P.~Pietzonka, F.~Ritort and U.~Seifert,
\newblock \emph{Finite-time generalization of the thermodynamic uncertainty
  relation},
\newblock Phys. Rev. E \textbf{96}, 012101 (2017),
\newblock \doi{10.1103/PhysRevE.96.012101}.

\bibitem{PhysRevE.96.020103}
J.~M. Horowitz and T.~R. Gingrich,
\newblock \emph{Proof of the finite-time thermodynamic uncertainty relation for
  steady-state currents},
\newblock Phys. Rev. E \textbf{96}, 020103 (2017),
\newblock \doi{10.1103/PhysRevE.96.020103}.

\bibitem{roldan2015decision}
{\'E}.~Rold{\'a}n, I.~Neri, M.~D{\"o}rpinghaus, H.~Meyr and F.~J{\"u}licher,
\newblock \emph{Decision making in the arrow of time},
\newblock Physical review letters \textbf{115}(25), 250602 (2015),
\newblock \doi{10.1103/PhysRevLett.115.250602}.

\bibitem{saito2016waiting}
K.~Saito and A.~Dhar,
\newblock \emph{Waiting for rare entropic fluctuations},
\newblock EPL (Europhysics Letters) \textbf{114}(5), 50004 (2016),
\newblock \doi{10.1209/0295-5075/114/50004}.

\bibitem{PhysRevE.95.032134}
J.~P. Garrahan,
\newblock \emph{Simple bounds on fluctuations and uncertainty relations for
  first-passage times of counting observables},
\newblock Phys. Rev. E \textbf{95}, 032134 (2017),
\newblock \doi{10.1103/PhysRevE.95.032134}.

\bibitem{gringich2017bis}
T.~R. Gingrich and J.~M. Horowitz,
\newblock \emph{Fundamental bounds on first passage time fluctuations for
  currents},
\newblock Phys. Rev. Lett. \textbf{119}, 170601 (2017),
\newblock \doi{10.1103/PhysRevLett.119.170601}.

\bibitem{PhysRevE.103.L050103}
K.~Hiura and S.-i. Sasa,
\newblock \emph{Kinetic uncertainty relation on first-passage time for
  accumulated current},
\newblock Phys. Rev. E \textbf{103}, L050103 (2021),
\newblock \doi{10.1103/PhysRevE.103.L050103}.

\bibitem{PhysRevResearch.3.L032034}
A.~Pal, S.~Reuveni and S.~Rahav,
\newblock \emph{Thermodynamic uncertainty relation for first-passage times on
  markov chains},
\newblock Phys. Rev. Research \textbf{3}, L032034 (2021),
\newblock \doi{10.1103/PhysRevResearch.3.L032034}.

\bibitem{neri2021universal}
I.~Neri,
\newblock \emph{Universal tradeoff relation between speed, uncertainty, and
  dissipation in nonequilibrium stationary states},
\newblock SciPost Phys. \textbf{12}(139) (2020),
\newblock \doi{10.21468/SciPostPhys.12.4.139}.

\bibitem{wampler2021skewness}
T.~Wampler and A.~C. Barato,
\newblock \emph{Skewness and kurtosis in stochastic thermodynamics},
\newblock Journal of Physics A: Mathematical and Theoretical \textbf{55}(1),
  014002 (2021),
\newblock \doi{10.1088/1751-8121/ac3b0c}.

\bibitem{PhysRevE.105.044127}
Y.~Hasegawa,
\newblock \emph{Thermodynamic uncertainty relation for quantum first-passage
  processes},
\newblock Phys. Rev. E \textbf{105}, 044127 (2022),
\newblock \doi{10.1103/PhysRevE.105.044127}.

\bibitem{neri2}
I.~Neri,
\newblock \emph{Estimating entropy production rates with first-passage
  processes},
\newblock J. Phys. A: Math. Theor. \textbf{55}, 304005 (2022),
\newblock \doi{10.1088/1751-8121/ac736b}.

\bibitem{hartich2019extreme}
D.~Hartich and A.~Godec,
\newblock \emph{Extreme value statistics of ergodic markov processes from first
  passage times in the large deviation limit},
\newblock Journal of Physics A: Mathematical and Theoretical \textbf{52}(24),
  244001 (2019),
\newblock \doi{10.1088/1751-8121/ab1eca}.

\bibitem{majumdar2020extreme}
S.~N. Majumdar, A.~Pal and G.~Schehr,
\newblock \emph{Extreme value statistics of correlated random variables: a
  pedagogical review},
\newblock Physics Reports \textbf{840}, 1 (2020),
\newblock \doi{10.1016/j.physrep.2019.10.005}.

\bibitem{chetrite2011two}
R.~Chetrite and S.~Gupta,
\newblock \emph{Two refreshing views of fluctuation theorems through kinematics
  elements and exponential martingale},
\newblock Journal of Statistical Physics \textbf{143}(3), 543 (2011),
\newblock \doi{10.1007/s10955-011-0184-0}.

\bibitem{neri2017statistics}
I.~Neri, {\'E}.~Rold{\'a}n and F.~J{\"u}licher,
\newblock \emph{Statistics of infima and stopping times of entropy production
  and applications to active molecular processes},
\newblock Physical Review X \textbf{7}(1), 011019 (2017),
\newblock \doi{10.1103/PhysRevX.7.011019}.

\bibitem{neri2019integral}
I.~Neri, {\'E}.~Rold{\'a}n, S.~Pigolotti and F.~J{\"u}licher,
\newblock \emph{Integral fluctuation relations for entropy production at
  stopping times},
\newblock Journal of Statistical Mechanics: Theory and Experiment
  \textbf{2019}(10), 104006 (2019),
\newblock \doi{10.1088/1742-5468/ab40a0}.

\bibitem{PhysRevE.105.024112}
G.~Manzano and E.~Rold\'an,
\newblock \emph{Survival and extreme statistics of work, heat, and entropy
  production in steady-state heat engines},
\newblock Phys. Rev. E \textbf{105}, 024112 (2022),
\newblock \doi{10.1103/PhysRevE.105.024112}.

\bibitem{singh2019extreme}
S.~Singh, {\'E}.~Rold{\'a}n, I.~Neri, I.~M. Khaymovich, D.~S. Golubev, V.~F.
  Maisi, J.~T. Peltonen, F.~J{\"u}licher and J.~P. Pekola,
\newblock \emph{Extreme reductions of entropy in an electronic double dot},
\newblock Physical Review B \textbf{99}(11), 115422 (2019),
\newblock \doi{10.1103/PhysRevB.99.115422}.

\bibitem{PhysRevE.102.062127}
K.~Cheng, J.-Q. Dong, W.-H. Han, F.~Liu and L.~Huang,
\newblock \emph{Infima statistics of entropy production in an underdamped
  brownian motor},
\newblock Phys. Rev. E \textbf{102}, 062127 (2020),
\newblock \doi{10.1103/PhysRevE.102.062127}.

\bibitem{liptser1977statistics}
R.~Liptser and A.~N. Shiryaev,
\newblock \emph{Statistics of random processes:I. General theory},
\newblock Springer Science \& Business Media, 2nd edn. (2013).

\bibitem{PhysRevE.101.022129}
Y.-J. Yang and H.~Qian,
\newblock \emph{Unified formalism for entropy production and fluctuation
  relations},
\newblock Phys. Rev. E \textbf{101}, 022129 (2020),
\newblock \doi{10.1103/PhysRevE.101.022129}.

\bibitem{chetrite2019martingale}
R.~Ch{\'e}trite, S.~Gupta, I.~Neri and {\'E}.~Rold{\'a}n,
\newblock \emph{Martingale theory for housekeeping heat},
\newblock EPL (Europhysics Letters) \textbf{124}(6), 60006 (2019),
\newblock \doi{10.1209/0295-5075/124/60006}.

\bibitem{roldan2022martingales}
{\'E}.~Rold{\'a}n, I.~Neri, R.~Chetrite, S.~Gupta, S.~Pigolotti,
  F.~J{\"u}licher and K.~Sekimoto,
\newblock \emph{Martingales for physicists},
\newblock arXiv preprint arXiv:2210.09983  (2022).

\bibitem{polettini2017effective}
M.~Polettini and M.~Esposito,
\newblock \emph{Effective thermodynamics for a marginal observer},
\newblock Physical review letters \textbf{119}(24), 240601 (2017),
\newblock \doi{10.1103/PhysRevLett.119.240601}.

\bibitem{polettini2019effective}
M.~Polettini and M.~Esposito,
\newblock \emph{Effective fluctuation and response theory},
\newblock Journal of Statistical Physics \textbf{176}(1), 94 (2019),
\newblock \doi{10.1007/s10955-019-02291-7}.

\bibitem{PhysRevLett.117.180601}
B.~Altaner, M.~Polettini and M.~Esposito,
\newblock \emph{Fluctuation-dissipation relations far from equilibrium},
\newblock Phys. Rev. Lett. \textbf{117}, 180601 (2016),
\newblock \doi{10.1103/PhysRevLett.117.180601}.

\bibitem{bisker2017hierarchical}
G.~Bisker, M.~Polettini, T.~R. Gingrich and J.~M. Horowitz,
\newblock \emph{Hierarchical bounds on entropy production inferred from partial
  information},
\newblock Journal of Statistical Mechanics: Theory and Experiment
  \textbf{2017}(9), 093210 (2017),
\newblock \doi{10.1088/1742-5468/aa8c0d}.

\bibitem{Liepelt}
S.~Liepelt and R.~Lipowsky,
\newblock \emph{Kinesin's network of chemomechanical motor cycles},
\newblock Phys. Rev. Lett. \textbf{98}, 258102 (2007),
\newblock \doi{10.1103/PhysRevLett.98.258102}.

\bibitem{hwang2018energetic}
W.~Hwang and C.~Hyeon,
\newblock \emph{Energetic costs, precision, and transport efficiency of
  molecular motors},
\newblock The journal of physical chemistry letters \textbf{9}(3), 513 (2018),
\newblock \doi{10.1021/acs.jpclett.7b03197}.

\bibitem{hwang2019correction}
W.~Hwang and C.~Hyeon,
\newblock \emph{Correction to “energetic costs, precision, and transport
  efficiency of molecular motors”},
\newblock The Journal of Physical Chemistry Letters \textbf{10}(12), 3472
  (2019),
\newblock \doi{10.1021/acs.jpclett.9b01630}.

\bibitem{Goethe}
J.~W.~v. Goethe,
\newblock \emph{Elective Affinities A Novel},
\newblock OUP Oxford,
\newblock Translated by David Constantine in 2008 (1809).

\bibitem{guillet2020extreme}
A.~Guillet, E.~Rold{\'a}n and F.~J{\"u}licher,
\newblock \emph{Extreme-value statistics of stochastic transport processes},
\newblock New Journal of Physics \textbf{22}(12), 123038 (2020),
\newblock \doi{10.1088/1367-2630/abcf69}.

\bibitem{bremaud2013markov}
P.~Br{\'e}maud,
\newblock \emph{Markov chains: Gibbs fields, Monte Carlo simulation, and
  queues}, vol.~31,
\newblock Springer-Verlag, 1st edn. (1999).

\bibitem{seifert2012stochastic}
U.~Seifert,
\newblock \emph{Stochastic thermodynamics, fluctuation theorems and molecular
  machines},
\newblock Reports on progress in physics \textbf{75}(12), 126001 (2012),
\newblock \doi{10.1088/0034-4885/75/12/126001}.

\bibitem{sekimoto2010stochastic}
K.~Sekimoto,
\newblock \emph{Stochastic energetics}, vol. 799,
\newblock Springer (2010).

\bibitem{10.21468/SciPostPhysLectNotes.32}
C.~Maes,
\newblock \emph{{Local detailed balance}},
\newblock SciPost Phys. Lect. Notes p.~32 (2021),
\newblock \doi{10.21468/SciPostPhysLectNotes.32}.

\bibitem{prigogine1978time}
I.~Prigogine,
\newblock \emph{Time, structure, and fluctuations},
\newblock Science \textbf{201}(4358), 777 (1978),
\newblock \doi{10.1126/science.201.4358.777}.

\bibitem{ge2021martingale}
H.~Ge, C.~Jia and X.~Jin,
\newblock \emph{Martingale structure for general thermodynamic functionals of
  diffusion processes under second-order averaging},
\newblock Journal of Statistical Physics \textbf{184}(2), 1 (2021),
\newblock \doi{10.1007/s10955-021-02798-y}.

\bibitem{gingrich2017inferring}
T.~R. Gingrich, G.~M. Rotskoff and J.~M. Horowitz,
\newblock \emph{Inferring dissipation from current fluctuations},
\newblock Journal of Physics A: Mathematical and Theoretical \textbf{50}(18),
  184004 (2017),
\newblock \doi{10.1088/1751-8121/aa672f}.

\bibitem{gomez2008lower}
A.~Gomez-Marin, J.~M.~R. Parrondo and C.~Van~den Broeck,
\newblock \emph{Lower bounds on dissipation upon coarse graining},
\newblock Phys. Rev. E \textbf{78}, 011107 (2008),
\newblock \doi{10.1103/PhysRevE.78.011107}.

\bibitem{roldan2010estimating}
E.~Rold\'an and J.~M.~R. Parrondo,
\newblock \emph{Estimating dissipation from single stationary trajectories},
\newblock Phys. Rev. Lett. \textbf{105}, 150607 (2010),
\newblock \doi{10.1103/PhysRevLett.105.150607}.

\bibitem{uhl2018fluctuations}
M.~Uhl, P.~Pietzonka and U.~Seifert,
\newblock \emph{Fluctuations of apparent entropy production in networks with
  hidden slow degrees of freedom},
\newblock Journal of Statistical Mechanics: Theory and Experiment
  \textbf{2018}(2), 023203 (2018),
\newblock \doi{10.1088/1742-5468/aaa78b}.

\bibitem{martinez2019inferring}
I.~A. Mart{\'\i}nez, G.~Bisker, J.~M. Horowitz and J.~M. Parrondo,
\newblock \emph{Inferring broken detailed balance in the absence of observable
  currents},
\newblock Nature communications \textbf{10}(1), 1 (2019),
\newblock \doi{10.1038/s41467-019-11051-w}.

\bibitem{harunari2022learn}
P.~E. Harunari, A.~Dutta, M.~Polettini and E.~Rold\'an,
\newblock \emph{What to learn from a few visible transitions' statistics?},
\newblock Phys. Rev. X \textbf{12}, 041026 (2022),
\newblock \doi{10.1103/PhysRevX.12.041026}.

\bibitem{van2022thermodynamic}
J.~van~der Meer, B.~Ertel and U.~Seifert,
\newblock \emph{Thermodynamic inference in partially accessible markov
  networks: A unifying perspective from transition-based waiting time
  distributions},
\newblock Phys. Rev. X \textbf{12}, 031025 (2022),
\newblock \doi{10.1103/PhysRevX.12.031025}.

\bibitem{PhysRevLett.124.120603}
S.~K. Manikandan, D.~Gupta and S.~Krishnamurthy,
\newblock \emph{Inferring entropy production from short experiments},
\newblock Phys. Rev. Lett. \textbf{124}, 120603 (2020),
\newblock \doi{10.1103/PhysRevLett.124.120603}.

\bibitem{PhysRevLett.122.220602}
G.~Manzano, R.~Fazio and E.~Rold\'an,
\newblock \emph{Quantum martingale theory and entropy production},
\newblock Phys. Rev. Lett. \textbf{122}, 220602 (2019),
\newblock \doi{10.1103/PhysRevLett.122.220602}.

\bibitem{PhysRevLett.124.040601}
I.~Neri,
\newblock \emph{Second law of thermodynamics at stopping times},
\newblock Phys. Rev. Lett. \textbf{124}, 040601 (2020),
\newblock \doi{10.1103/PhysRevLett.124.040601}.

\bibitem{PhysRevLett.126.080603}
G.~Manzano, D.~Subero, O.~Maillet, R.~Fazio, J.~P. Pekola and E.~Rold\'an,
\newblock \emph{Thermodynamics of gambling demons},
\newblock Phys. Rev. Lett. \textbf{126}, 080603 (2021),
\newblock \doi{10.1103/PhysRevLett.126.080603}.

\bibitem{manzano2022quantum}
G.~Manzano and R.~Zambrini,
\newblock \emph{Quantum thermodynamics under continuous monitoring: a general
  framework},
\newblock AVS Quantum Science \textbf{4}(2), 025302 (2022),
\newblock \doi{10.1116/5.0079886}.

\bibitem{PhysRevE.101.062139}
C.~Moslonka and K.~Sekimoto,
\newblock \emph{Memory through a hidden martingale process in progressive
  quenching},
\newblock Phys. Rev. E \textbf{101}, 062139 (2020),
\newblock \doi{10.1103/PhysRevE.101.062139}.

\bibitem{PhysRevE.105.044146}
C.~Moslonka and K.~Sekimoto,
\newblock \emph{Martingale-induced local invariance in progressive quenching},
\newblock Phys. Rev. E \textbf{105}, 044146 (2022),
\newblock \doi{10.1103/PhysRevE.105.044146}.

\bibitem{williams1991probability}
D.~Williams,
\newblock \emph{Probability with martingales},
\newblock Cambridge university press (1991).

\end{thebibliography}

\nolinenumbers

\end{document}